\documentclass[letterpaper, 11pt]{article}

\usepackage{amsmath,amssymb}
\usepackage{bm,bbm}
\usepackage{graphicx}
\usepackage{cite}
\usepackage{pstricks}
\usepackage{color}
\usepackage{subfigure}

\usepackage[normalem]{ulem}

\usepackage[letterpaper, margin=1in]{geometry}
\numberwithin{equation}{section}

\newcommand{\eq}[1]{(\ref{e.#1})}
\newcommand{\eql}[1]{\label{e.#1}}
\newcommand{\Fig}[1]{Fig.~\ref{f.#1}}

\newcommand{\Sec}[1]{Section~\ref{s.#1}}
\newcommand{\Secs}[1]{Sections~\ref{s.#1}}

\newcommand{\beq}{\begin{equation}\begin{aligned}}
\newcommand{\eeq}{\end{aligned}\end{equation}}
\newcommand{\beqa}[1]{\begin{equation}\begin{alignedat}{#1}}
\newcommand{\eeqa}{\end{alignedat}\end{equation}}

\newcommand{\fr}[2]{\dfrac{#1}{#2}}
\newcommand{\wba}[1]{\overline{#1}}
\newcommand{\wtd}[1]{\widetilde{#1}}

\newcommand{\diag}{\mathrm{diag}\,}

\newcommand{\lt}{\!\left}
\newcommand{\rt}{\right}

\newcommand{\lr}{\leftrightarrow}

\newcommand{\too}{\longrightarrow}

\newcommand{\dt}{\!\cdot\!}

\newcommand{\del}{\partial}
\newcommand{\dd}{\mathrm{d}}
\newcommand{\DD}{\mathrm{D}}

\newcommand{\I}{\mathrm{i}}
\newcommand{\e}{\mathrm{e}}
\newcommand{\C}{\mathrm{c}}

\newcommand{\al}{\alpha}
\newcommand{\be}{\beta}
\newcommand{\ga}{\gamma}
\newcommand{\Ga}{\Gamma}
\newcommand{\de}{\delta}
\newcommand{\ep}{\epsilon}
\newcommand{\vep}{\varepsilon}
\newcommand{\ka}{\kappa}
\newcommand{\la}{\lambda}
\newcommand{\La}{\Lambda}
\newcommand{\om}{\omega}
\newcommand{\sg}{\sigma}

\newcommand{\Sg}{\Sigma}
\newcommand{\vphi}{\varphi}

\newcommand{\cA}{{\mathcal A}}

\newcommand{\cL}{{\mathcal L}}
\newcommand{\cM}{{\mathcal M}}
\newcommand{\cO}{{\mathcal O}}
\newcommand{\cP}{{\mathcal P}}

\newcommand{\cS}{{\mathcal S}}

\newcommand{\tev}{\>\text{TeV}}

\newcommand{\Mpl}{M_\text{Pl}}

\newcommand{\SU}{\mathrm{SU}}

\newcommand{\U}{\mathrm{U}}

\newcommand{\PS}{{\phantom.}}
\newcommand{\PD}{{\phantom\dag}}
\newcommand{\ds}[1]{\displaystyle{#1}}

\newcommand{\ps}{{\text{\tiny$+$}}}
\newcommand{\ms}{{\text{\tiny$-$}}}
\newcommand{\pp}{{\text{\tiny$\perp$}}}
\newcommand{\pms}{{\text{\tiny$\pm$}}}
\newcommand{\mps}{{\text{\tiny$\mp$}}}

\newcommand{\s}{\mathrm{s}}

\begin{document}
\begin{titlepage}

~\\
\vskip 1in

\begin{center}

{\Huge\bf Soft collinear effective theory for gravity}

\vskip .5in

{\Large {\bf Takemichi Okui} and {\bf Arash Yunesi}}

\vskip .5in 

{\it Department of Physics, Florida State University,\\
77 Chieftain Way, Tallahassee, FL 32306, USA}

\vskip 1in

\abstract{We present how to construct a Soft Collinear Effective Theory (SCET) for gravity at the leading and next-to-leading powers from the ground up. The soft graviton theorem and decoupling of collinear gravitons at the leading power are manifest from the outset in the effective symmetries of the theory. At the next-to-leading power, certain simple structures of amplitudes, which are completely obscure in Feynman diagrams of the full theory, are also revealed, which greatly simplifies calculations. The effective lagrangian is highly constrained by effectively multiple copies of diffeomorphism invariance that are inevitably present in gravity SCET due to mode separation, an essential ingredient of any SCET\@. Further explorations of effective theories of gravity with mode separation may shed light on lagrangian-level understandings of some of the surprising properties of gravitational scattering amplitudes. A gravity SCET with an appropriate inclusion of Glauber modes may serve as a powerful tool for studying gravitational scattering in the Regge limit.}

\end{center}

\end{titlepage}

\tableofcontents
\vskip .5in

\section{Introduction}
\label{s.intro}
Quantum gravity is a fascinating frontier of theoretical physics as it is the only known place where two otherwise perfect theoretical structures---general relativity and quantum mechanics---dramatically clash with each other.
Even if we limit ourselves to the study of the ``known'' effective field theory (EFT) of gravity beneath the Planck scale, its scattering amplitudes display many surprising properties that are completely invisible at the lagrangian level such as 
the well-known soft theorem and decoupling of collinear graviton (in the Eikonal approximation~\cite{Weinberg:1965nx} and in a full diagrammatic analysis~\cite{Akhoury:2011kq}), infinite dimensional symmetries~\cite{Strominger:2013jfa, He:2014laa, Kapec:2014opa}, and ``gravity $=$ gauge$^2$''~\cite{Kawai:1985xq, Bern:2008qj}. 
It is possible that some of those amazing properties of the amplitudes may be already revealed at the lagrangian level if we perform the path integral partially instead of going all the way to the amplitudes.
Therefore, we are motivated to construct a ``more effective'' theory of gravity by integrating more modes out of the standard EFT of gravity.

In this paper, as an example of such theories,
we develop a Soft Collinear Effective Theory (SCET) for gravity
at the leading and next-to-leading powers of $\la$.
SCET is an EFT originally developed in~\cite{Bauer:2000ew, Bauer:2000yr, Bauer:2001ct, Bauer:2001yt, Bauer:2002nz} in the context of QCD for systematically and efficiently calculating amplitudes for clusters of highly collimated energetic particles with soft radiations,
where $\la$ is a small parameter used to characterize such region of the phase space and the lagrangian of a SCET is an expansion in powers of $\la$.
Ref.~\cite{Beneke:2012xa} made an early attempt to construct a gravity SCET at the leading power.
However, our gravity SCET disagrees with theirs, even at the leading power, in very fundamental ways such as the symmetry structures and the forms of gravitational Wilson lines. We will discuss the differences in \Sec{comparison}.

As SCET is largely an unfamiliar subject outside the QCD community,
and also to avoid being misguided by some structures of QCD SCET that are not shared by gravity, 
we will develop a gravity SCET from the ground up, 
reviewing general concepts of SCET and introducing some specifics of gravity SCET in \Sec{basics}, moving on to the constructions of the leading power and next-to-leading power SCET lagrangians in \Secs{leading_order} and~\ref{s.order-lambda}.
In Appendices~\ref{a.Phi4}--\ref{a.ssff} we give explicit examples of full-theory amplitudes expanded to the next-to-leading power and show how they match the structures predicted by the gravity SCET\@.

We will find that the structure of the gravity SCET lagrangian dictated by the effective symmetries and power counting rules indeed reveals many properties of gravity amplitudes that are completely obscure in the full theory (i.e., the ``full EFT'' of gravity valid beneath the Planck scale), without recourse to any actual calculations.  
For example, we will see that the gravitational soft theorem and absence of collinear IR divergences at the leading power are manifestly dictated from the outset by the symmetries of the effective lagrangian.
Soft IR divergences may be present at the leading power but their sources will be explicitly isolated in the effective lagrangian.
At the next-to-leading power, the structure of the gravity SCET lagrangian immediately translates to certain structures in the amplitudes.
As illustrated by the explicit examples in Appendices~\ref{a.Phi4}--\ref{a.ssff},
this can greatly simplify the calculations of amplitudes, where a lengthy full-theory calculation with laborious expansions in $\la$ and tricky cancellations of many terms can be reproduced by short scribbles in the EFT to calculate the few terms just enough for determining the entire amplitudes.

We will also see other interesting structures of gravity SCET that may be relevant for deeper understanding of gravity.
For example, in a gravity SCET that describes $N$ distinct collinear directions,
the diffeomorphism (diff) invariance of the full general relativity turns into $N$ copies of diff invariance ``effectively'' (the precise meaning of which will be explained in \Sec{fact_symm}),
as in a QCD SCET with $N$ collinear sectors ``effectively'' has $N$ copies of $\SU(3)_\text{c}$ gauge invariance. This suggests that the original full-theory $S$-matrix (which ``contains'' all SCETs with different values of $N$) must have an infinite number of effective diff invariances in some way.
It would be interesting to explore connections between this and the infinite dimensional symmetries of gravitational scattering amplitudes discovered recently~\cite{Strominger:2013jfa, He:2014laa, Kapec:2014opa}.
We will also find that 
the $N$ copies of diff invariance in gravity SCET lead to specific variations of nonlocal ``dressing'' of operators discussed in~\cite{Donnelly:2015hta, Donnelly:2016rvo}.
We will be led to our specific forms of nonlocal dressing as a consequence of effective gauge symmetries of SCET at long distances, while Refs.~\cite{Donnelly:2015hta, Donnelly:2016rvo} arrived at a general notion of dressing by trying to answer the question of what might be the fundamental observables in the yet-to-be-found ultimate quantum theory of gravity.

\section{Building up gravity SCET}
\label{s.basics}
In this section we build gravity SCET from the ground up starting from fundamental principles of EFT\@.
Along the way, we review essential conceptual elements of SCET in order to make the paper accessible to the reader who is not familiar with ``modern'' effective field theories with a property called \emph{mode separation} (to be discussed below). Some examples of such EFTs are  Non-Relativistic QCD (NRQCD)~\cite{Caswell:1985ui, Luke:1999kz}, Non-Relativistic General Relativity (NRGR)~\cite{Goldberger:2004jt}, and SCET\@. 

\subsection{Fundamentals}

\subsubsection{The target phase space}
\label{s.target}
By design, an EFT is aimed only at a prescribed, limited region of the phase space characterized by a small parameter or parameters 
(e.g., ``all particles have energy much below $\ds{1\tev}$''). 
We therefore must begin by defining the target phase space of our gravity SCET.

Imagine a scattering process in 4d Minkowski spacetime.%
\footnote{\label{fn:metric}%
All spacetime indices in this paper will be raised/lowered/contracted via the Minkowski metric in the $+\!-\!--$ convention.}
Identify most energetic particles in the initial and final states
and cluster them into multiple \emph{collinear sectors,}
where a collinear sector is defined as a set of energetic particles (with energy of $\cO(Q)$) moving in similar directions (with angular spread of $\cO(\la) \ll 1$).
We require different collinear sectors to be well separated in direction.
If two sectors are too similar in direction, merge them into one sector by choosing a larger $\la$.
(If such merger leads to $\la$ of $\cO(1)$, our SCET is not an appropriate EFT for the process in question.)  
After thus identifying all collinear sectors,
we are left with non-energetic particles, which we will collectively refer to as \emph{the soft sector.} 
We focus on the processes where soft particles are around only because they are \emph{required by nature} to be around given the presence of the collinear sectors, 
not because \emph{we intended} to include them.
The energy scale of the soft sector will then turn out to be $\cO(\la^2 Q)$ (explained in \Sec{modeseparation}).
Our target phase space is thus defined by the number of collinear sectors $N$, the \emph{hard energy scale} $Q$, and the small parameter $\la \ll 1$.
The most important parameter is $\la$, which characterizes how well collimated the energetic particles are in each collinear sector.
Our SCET effective lagrangian will be an expansion in powers of $\la$.

It should be noted that each collinear sector in principle comes with its own $\la$ and $Q$.
To avoid an overly general presentation that beclouds main points, however,    
we assume a common $\la$ and a common $Q$ for all the collinear sectors.
We will scale them independently only when it is necessary or convenient to do so.

In order to focus on the properties of SCET that are solely associated with gravity, 
we make another simplifying assumption that a collinear splitting due to non-gravitational interactions such as QCD occurs with a splitting angle that is either much larger or much smaller than $\la$.
In the former case, we regard a non-gravitational ``collinear'' splitting as actually giving rise to two distinct collinear sectors.
For the latter, we view the stream of almost exactly collinear, non-gravitationally splitting particles as a single massless ``particle''. 
The former would be a two-stage EFT where the full theory is first matched to a well-established gauge-theory SCET with a large ``$\la$'', which is then matched to a gravity SCET developed in this paper.
The latter would be a two-stage EFT with the reversed order.

\subsubsection{Manifest power counting}
\label{s.manifest}
To have a systematic control of the special kinematics of its target phase space, 
an EFT must be equipped with well-defined rules for power-counting the small parameters that define the target phase space.
For a SCET, this means that each term in its effective lagrangian must scale with a definite power of $\la$ so that we can ignore higher order terms irrelevant for achieving the desired precision.
But this is not good enough.
If interaction terms in the lagrangian are allowed by the symmetries of an EFT and appear to be the largest contributions in terms of its power counting rules, 
they should not be shown to be actually absent or exhibit any ``unexpected'' systematic cancellations.
The seeming existence of such terms would indicate that we have a wrong EFT 
and the theory should be revised such that those ``largest'' terms would be manifestly absent from the outset by being forbidden by symmetry or deemed subleading by power counting.
To the best of our knowledge, our theory is the first formulation of gravity SCET that passes this test of manifest power counting (see the end of \Sec{summary:leading_SCET} for more on this point and also \Sec{comparison} for an earlier attempt on SCET for gravity \cite{Beneke:2012xa} where the symmetries and power counting rules allow us to write down terms that can actually be systematically removed).

Restricting the phase space to the target phase space requires that the momentum space of virtual particles should be likewise restricted in an EFT\@.
This is to ensure that power counting be manifestly compatible with unitarity,
by having the momentum integrations along a cut scale in the same way as the corresponding phase space integrations in the optical theorem.
So, in our SCET, only the collinear and soft momentum modes are allowed even for virtual particles.
As usual in an EFT, virtual contributions from ``missing'' modes outside the target space can be systematically restored by including all possible effective interactions allowed by symmetry in the effective lagrangian and suitably adjusting their coefficients.  
There are an infinite number of such interactions, which is why we need well-defined power counting rules that allow us to truncate the effective lagrangian in a controlled way.

\subsubsection{Other underlying assumptions}
We assume that all particles have no or negligible mass compared to the soft energy scale, $\la^2 Q$, which is the lowest energy scale in the effective theory.
If this assumption is violated, our SCET is not an appropriate description of the physics except in one situation:
if all massive particles are much heavier than $Q$, 
we can just integrate them out to obtain an EFT expanded in inverse powers of those heavy masses.
We can then regard this EFT as the ``full'' theory that our SCET provides an effective description of. 
Ultimately, the presence of gravity means that the full theory is necessarily an EFT expanded in powers of $Q / \Mpl$.

\subsection{The collinear lightcone coordinates}
Let's introduce some notations for describing our target phase space in a manner convenient for power counting.
We index the collinear sectors by $i =1,2, \ldots, N$.
For each $i$,
we introduce a pair of null 4-vectors $n_{\ps_i}$ and $n_{\ms_i}$ 
satisfying
\beq 
n_{\ps_i} \dt\, n_{\ps_i} = n_{\ms_i} \dt\, n_{\ms_i} = 0
\,,\quad
n_{\ps_i} \dt\, n_{\ms_i} = 1
\,,\eql{n:conditions}
\eeq
where the spatial part, $\vec{n}_{\ps_i}$, of $n_{\ps_i}$ is taken to be in the direction of the $i$-th collinear sector, modulo an angular ambiguity of $\cO(\la)$.%
\footnote{No summation is implied for the repeated sector index $i$ here and throughout the paper.
A summation over collinear sectors will always be indicated explicitly.}%
${}^{,}$%
\footnote{Our normalization convention differs from the standard normalization in the SCET literature, $\ds{n_{\ps_i} \dt\, n_{\ms_i}} = 2$, to prevent twos and halves from appearing when raising or lowering $\pm$ indices.}
It is convenient to also define 4-vectors $n^{\ps_i}$ and $n^{\ms_i}$ as 
\beq
n^{\ps_i} \equiv n_{\ms_i}
\,,\quad
n^{\ms_i} \equiv n_{\ps_i}
\eeq
so that we have
\beq
n^{\pms_i} \dt\, n_{\pms_i} = 1
\,,\quad
n^{\pms_i} \dt\, n_{\mps_i} = 0
\,.
\eeq
In this basis, an arbitrary 4-vector $a$ can be written as
\beq
a^\mu 
&= a^{\ps_i} n_{\ps_i}^\mu + a^{\ms_i} n_{\ms_i}^\mu + a_{\pp_i}^\mu
\\
&= a_{\ps_i} n^{\ps_i}{}^\mu + a_{\ms_i} n^{\ms_i}{}^\mu + a_{\pp_i}^\mu
\eeq
with
\beq
a^{\ps_i} \equiv n^{\ps_i} \dt a
\,,\quad
a^{\ms_i} \equiv n^{\ms_i} \dt a
\,,\quad
n^{\pms_i} \dt a_{\pp_i} = 0
\eeq
and
\beq
a_{\ps_i} \equiv n_{\ps_i} \dt a = a^{\ms_i}
\,,\quad
a_{\ms_i} \equiv n_{\ms_i} \dt a = a^{\ps_i}
\,.
\eeq
Then, for arbitrary 4-vectors $a$ and $b$, we have 
\beq
a \dt b 
&= a_{\ps_i} b^{\ps_i} + a_{\ms_i} b^{\ms_i} + a_{\pp_i} \dt b_{\pp_i}
\\
&= a^{\ms_i} b^{\ps_i} + a^{\ps_i} b^{\ms_i} + a_{\pp_i} \dt b_{\pp_i}
\\
&= a_{\ps_i} b_{\ms_i} + a_{\ms_i} b_{\ps_i} + a_{\pp_i} \dt b_{\pp_i}
\,.
\eeq
%

\subsection{Mode separation and scaling}
\label{s.modeseparation}
In an EFT, manifest power counting often requires \emph{mode separation,} i.e.,
a further division of the target phase space into subregions,
if modes in different subregions are found to scale differently in terms of the small parameters that define the target phase space.
Mode separation is arguably \emph{the} feature that distinguishes ``modern'' EFTs like SCET from the classic Wilsonian EFTs.
In the latter, there is only one type of momentum modes, which scale as $\La^{-1}$ with the cutoff $\La$.
In more general EFTs, mode separation is often necessary to achieve manifest power counting. 
Below, we describe the momentum modes and their scalings in our target phase space.

\subsubsection{The collinear momentum scaling}
Let's begin with the $\la$ scaling of momenta in the collinear sector along $\vec{n}_{\ps_i}$ of a given arbitrary $i$, or \emph{the $n_i$-collinear sector} for short.
Let $p$ be the 4-momentum of an $n_i$-collinear particle that is either on-shell or nearly on-shell due to emissions/absorptions of soft particles.
By definition, the particle is moving approximately in the $\vec{n}_{\ps_i}$ direction, carrying a large energy of $\cO(Q)$.
We thus have $p^{\ps_i} \sim Q$.
Next, again by definition, a typical angle between $\vec{n}_{\ps_i}$ and the exact direction of $\vec{p}$ is $\cO(\la)$.
Hence, $p_{\pp_i} \sim \la Q$.
Finally, 
in order for $p$ to be (nearly) on-shell, 
the $p^{\ps_i} p^{\ms_i}$ and $\ds{p_{\pp_i} \!\dt p_{\pp_i}}$ terms in $p^2 = 2 p^{\ps_i} p^{\ms_i} + \ds{p_{\pp_i} \!\dt p_{\pp_i}}$ must be of the same order in $\la$ so that they could add up to zero. 
Hence, $p^{\ms_i} \sim \ds{p_{\pp_i} \!\dt p_{\pp_i}} / p^{\ps_i} \sim \la^2 Q$. 

To summarize, an $n_i$-collinear momentum $p$ scales as
$(p^{\ps_i}, p^{\ms_i}, p_{\pp_i}) \sim Q(1, \la^2, \la)$,
which we simply write as $p \sim (1, \la^2, \la)_i$.
This implies $p^2 \sim \la^2 Q^2$, which we simply write as $p^2 \sim \la^2$.

\subsubsection{The cross-collinear scaling and reparametrization invariance (RPI)}
\label{s.cross-collinear-RPI}
Let's now consider the case in which we pick one momentum from the $n_i$-collinear sector and another from the $n_j$-collinear sector with $i \neq j$.
Since different collinear sectors in our target phase space are well separated in directions from each other,
we have
\beq
n_{\pms_i} \dt\, n_{\pms_j} \sim n_{\pms_i} \dt\, n_{\mps_j} \sim \la^0
\qquad (i \neq j)
\,,\eql{cross-collinear}
\eeq
which we dub \emph{the cross-collinear scaling.}
In other words, an $n_i$-collinear momentum $p_i$ scales as $p_i \sim (1,1,1)_j$ for all $j \neq i$.

We must be careful when $\vec{n}_{\ps_i}$ and $\vec{n}_{\ps_j}$ point back-to-back.
For example, for $\vec{n}_{\ps_i} = (0, 0, 1)$ and $\vec{n}_{\ps_j} = (0, 0, -1)$,
we might be tempted to choose $n_{\pms_i} \propto (1, 0, 0, \pm 1)$ and $n_{\pms_j} \propto (1, 0, 0, \mp 1)$. 
But this would lead to $\ds{n_{\ps_i} \dt\, n_{\ms_j}} = 0$, violating the cross-collinear scaling~\eq{cross-collinear}.
Notice, however, that the conditions~\eq{n:conditions} do not determine $n_{\ms_i}$ uniquely from a given $n_{\ps_i}$.
There is also freedom even in the choice of $n_{\ps_i}$, 
due to the $\cO(\la)$ ambiguity in the direction of $\vec{n}_{\ps_i}$ and the arbitrariness in the normalization of $n_{\ps_i}$.
These ambiguities are fundamental redundancies in SCET and a SCET lagrangian must be invariant under all possible redefinitions of $n_{\pms_i}$ that preserve the conditions~\eq{n:conditions} and the scaling $p \sim (1, \la^2, \la)_i$~\cite{Chay:2002vy, Manohar:2002fd}.
This property is called reparametrization invariance (RPI)
and it should be noted that each collinear sector comes with its own RPI\@.
So, a more precise statement is that
the cross-collinear scaling law~\eq{cross-collinear} holds for \emph{generic} choices of $n_{\pms_i}$ and $n_{\pms_j}$, 
even when $\vec{n}_{\ps_i}$ and $\vec{n}_{\ps_j}$ are back-to-back.
For example, 
for $\vec{n}_{\ps_i} = (0, 0, 1)$ and $\vec{n}_{\ps_j} = (0, 0, -1)$,
the cross-collinear scaling holds for 
the choice $n_{\ps_i} \propto (1, 0, 0, 1)$, $n_{\ms_i} \propto (5, 0, 4, 3)$, $n_{\ps_j} \propto (1, 0, 0, -1)$, $n_{\ms_j} \propto (5, 0, -4, -3)$.
In the remainder of the paper, we tacitly assume that a generic choice has been made such that the cross-collinear scaling law holds.

\subsubsection{The soft momentum scaling}
Let us now find the $\la$ scaling of soft momenta.
By definition, soft particles are much less energetic than the collinear particles,
with the energies low enough that soft particles can be added or removed without changing the existing kinematics of the collinear sectors that we have already defined.%
\footnote{If the attachment of a ``soft'' particle to an $n_i$-collinear line of some $i$ was found to knock the line's momentum out of the $(1, \la^2, \la)_i$ range, 
we should have actually defined a separate collinear sector for that ``soft'' particle in the first place with its own smaller value ``$Q$''.
Our simplifying assumption of universal $Q$ and $\la$, however, excludes such possibilities from consideration.}
They are also not preferentially associated with any particular direction.%
\footnote{If the soft particles as a whole had some directional preference, 
they should form their own collinear sector with a smaller $Q$ and a larger $\la$.  
Again, our simplifying assumption of universal $Q$ and $\la$ excludes such possibilities.}
Therefore, a soft 4-momentum must scale as $\sim (\la^a, \la^a, \la^a)_i$ for \emph{all} $i$ with some $a > 1$.

Now, imagine a generic Feynman diagram with $N$ collinear sectors and then attach a soft particle to an $n_i$-collinear line with momentum $p_\C$. 
The scaling of the $p_\C^{\ms_i}$ component (i.e., the second entry of $(1, \la^2, \la)_i$) tells us that $a$ must be $\geq 2$ in order for the attachment of the soft particle to preserve the $n_i$-collinearness of the $n_i$-collinear line.
Then, in order to minimize $\la$ suppressions from derivatives acting on the soft particle's field in the SCET lagrangian, 
we are \emph{interested} in the $a=2$ case to ensure that the largest soft contributions are taken into account by the effective lagrangian.%
\footnote{Our SCET thus belongs to the category of SCET often referred to as SCET$_\text{I}$ in the SCET literature, in which our soft particles tend to be called \emph{ultrasoft} particles.}

To summarize, a soft 4-momentum $p$ scales as $p \sim (\la^2, \la^2, \la^2)_i$ for all $i$, which we simply write as $p \sim (\la^2, \la^2, \la^2)$ without referring to any $i$.
Soft momenta have $p^2 \sim \la^4$.

\subsubsection{Comparison with Soft Graviton Effective Theory}
Before we move on to next step, we would like to compare our gravity SCET with SGET (Soft Graviton Effective Theory~\cite{Sundrum:2003jq, Okui:2005xh}), which was originally proposed by~\cite{Sundrum:2003jq} to demonstrate that a low cutoff ($\sim \mathrm{meV}$) that one might want to impose on the gravity sector to render the small cosmological constant natural is not necessarily in contradiction with the known high cutoff ($\gg \mathrm{TeV}$) of the standard-model matter sector. 
The difference between SGET and our gravity SCET---just like the difference between any two EFTs---is how their target phase spaces are prescribed.
In SGET, all graviton fields are soft, and matter fields only have soft fluctuations around a point on the mass shell, similarly to Heavy Quark Effective Theory (HQET)~\cite{Georgi:1990um} where all gluons are soft and quarks fluctuate only softly around a point on the mass shell.

\subsection{Mode-separating fields} 
As a first step toward manifest power counting,
we have separated momentum modes in our target phase space and classified them into distinct groups depending on how they scale with $\la$.
So, each particle species now comes with $N+1$ distinct propagators: 
the $n_i$-collinear propagators ($i=1, \ldots, N$) and the soft propagator.

This mode separation can be explicitly facilitated in a quantum field theory by introducing an independent interpolating field for each propagator type, 
for each particle species. 
That is, for each field $\Phi(x)$ of the full theory,%
\footnote{Recall that we have assumed all particles have no or negligible mass, 
so the particle content of the EFT agrees with that of the full theory.}
we introduce $N$ distinct \emph{collinear fields} $\Phi_i(x)$ ($i=1, \ldots, N$) and a \emph{soft field} $\Phi_\s(x)$, 
where $\Phi_i(x)$ and $\Phi_\s(x)$ only contain the $n_i$-collinear and soft Fourier modes, respectively.
That is, when $\del_\mu$ acts on $\Phi_\s$, it scales as $\del \sim (\la^2, \la^2, \la^2)$. When it acts on $\Phi_i$, it scales as $\del \sim (1, \la^2, \la)_i$.%
\footnote{\label{fn:spiritviolation}Since a soft momentum can be added to an $n_i$-collinear momentum without destroying the $n_i$-collinear scaling,
the $n_i$-collinear scaling is ambiguous up to a soft momentum by definition.
So, strictly speaking, the momentum modes in $\Phi_i$ actually scales like $\sim (1, \la^2, \la)_i + (\la^2, \la^2, \la^2)_i$,
which violates the spirit of manifest power counting. 
Strictly manifest power counting can be recovered at the expense of notational simplicity by introducing the so-called \emph{label momenta}   
as it was done in the original SCET papers~\cite{Bauer:2000ew, Bauer:2000yr, Bauer:2001ct, Bauer:2001yt, Bauer:2002nz}.
We opt for a notational simplicity and just \emph{remember} that $\Phi_i(x)$ also contains soft fluctuations, 
following the ``position space'' formulation of SCET~\cite{Beneke:2002ph, Beneke:2002ni}.
In particular,
since we will be concerning next-to-leading-power corrections later in this paper, we must watch out for the $\cO(\la^2)$ fluctuations in the $\perp$-components of collinear momenta.}
It should be emphasized that we are not increasing the degrees of freedom but merely giving different names to the different groups of Fourier modes of each $\Phi(x)$
in order to facilitate manifest power counting.

\subsection{Factorized effective gauge symmetry}
\label{s.fact_symm}
This is the most profound implication of mode separation and the heart of SCET as recognized in the original QCD SCET~\cite{Bauer:2001ct, Bauer:2001yt}, 
but it is a general observation that should also apply to gravity SCET\@.
Since the splitting of a full-theory field $\Phi$ into $\Phi_i$ ($i=1, \ldots, N$) and $\Phi_\s$ also equally applies to gauge fields (including the graviton field), 
each full-theory gauge symmetry $G$ (including the diffeomorphism invariance) splits into $N+1$ distinct gauge symmetries: 
the $n_i$-collinear gauge symmetries $G_i$ ($i=1, \ldots, N$) and the soft gauge symmetry $G_\s$.
An $n_i$-collinear gauge transformation $U_i(x) \in G_i$ should only contain $n_i$-collinear Fourier modes so that $U_i(x)$ maps the associated $n_i$-collinear gauge field to itself. 
Then, $U_i(x)$ also maps any other $\Phi_i$ to a $\Phi_i$.
Similarly, 
a soft gauge transformation $U_\s(x) \in G_\s$ only contains soft Fourier modes,
mapping a $\Phi_\s$ to a $\Phi_\s$ and also a $\Phi_i$ to a $\Phi_i$.
Therefore, a $\Phi_i$ is charged under both $G_i$ and $G_\s$,
while a $\Phi_s$ is charged under $G_\s$.
We must not charge a $\Phi_i$ under any $G_j$ with $j \neq i$,
because $U_j(x)$ with $j \neq i$ would not map a $\Phi_i$ to a $\Phi_i$, 
thereby jeopardizing mode separation in that mode separation would depend on the gauge choice.
Similarly, we must not charge a $\Phi_\s$ under $G_i$ with any $i$.
Such compatibility of gauge invariance and mode separation also requires us to forbid all the gauge transformations in $G$ that do not belong to any $G_i$ or $G_\s$.
Since there is still a one-to-one and onto correspondence between the allowed modes of gauge fields and those of gauge transformations, the EFT still possesses just enough gauge transformations to gauge away all unphysical polarizations of the gauge fields if we so wish.

How can we understand such splitting of $G \to G_{1, \ldots, N, \s}$ from a ``microscopic'' viewpoint? 
Imagine a set of all full-theory diagrams with the external lines in the target phase space. 
We can then ``derive'' the set of all EFT diagrams 
by identifying every full-theory propagator that do not belong to the target phase space 
and then shrinking all such propagators and tucking them into effective vertices.
Consider, for example, an $n_i$-collinear electron in the initial state and let it emit an $n_j$-collinear photon with $j \neq i$ in the full theory (i.e., QED\@).  
Then, the electron's propagator after the emission would necessarily be highly off-shell and out of the target phase space, 
so it must be shrunk and tucked into an effective vertex. 
Thus, in the EFT, such emission is described by some effective operator, not by a vertex from the covariant derivative in the $n_i$-collinear electron's kinetic term.
In contrast, if the $n_i$-collinear electron emits an $n_i$-collinear photon with the same $i$, the electron's propagator after the emission remains $n_i$-collinear, 
so such emission continues to be described by a vertex from the covariant kinetic term of the electron. 
Such heuristic considerations suggest that the only covariant derivatives that may act on the $n_i$-collinear electron are those associated with the $n_i$-collinear photon (with the same $i$) or the soft photon.
In other words, each collinear sector has its own gauge invariance associated with the collinear photon modes in that sector, 
and all sectors share a common gauge invariance associated with the soft photon.

To summarize conceptually, each full-theory gauge symmetry $G$ is reduced in the SCET to its subset as
\beq
G \>\too\> 
G_\s \ltimes (G_1 \times G_2 \times \cdots \times G_N)
\,,\eql{eff_gauge_symm}
\eeq
where the symbol $\ltimes$ indicates a semi-direct product as in the same convention as we would write $\text{Poincar\'e} =  \text{Lorentz} \ltimes \text{Translations}$.
The semi-direct product expresses the property that the generators of $G_{1, \ldots, N}$ are also charged under $G_\s$, just like the translation generators are also charged under the Lorentz group.
In our discussion above, the semi-direct product is manifested in the fact that a $\Phi_i$ is charged under both $G_i$ and $G_\s$, while a $\Phi_s$ is charged under $G_\s$.
In the theory of gravity in terms of the vierbein (or tetrad), the gravitational part $G_\text{grav}$ of $G$ is given by
\beq
G_\text{grav} = \text{(diffeomorphism)} \times \text{(local Lorentz group)}
\,,
\eeq
which we call ``diff$\times$Lorentz'' for short.  
Although \eq{eff_gauge_symm} is conceptually a \emph{reduction} of gauge symmetry $G$ to its subset corresponding to the collinear and soft modes,%
\footnote{This is why~\eq{eff_gauge_symm} with $G = G_\text{grav}$ does not imply multiple spacetimes.}
in practice it acts as if $G$ was enhanced to $N+1$ copies of $G$ when we try to constrain the structure of the effective lagrangian.
Essentially, this is the source of the power of SCET\@.
   
Unfortunately, the form of effective gauge symmetry as in~\eq{eff_gauge_symm} is not yet fully compatible with mode separation.
The problem is the $\ltimes$ symbol, that is, the fact that a $\Phi_i$ is charged under not only $G_i$ but also $G_\s$.
This means that a gauge covariant derivative acting on a $\Phi_i$ must contain both the $n_i$-collinear and soft gauge fields, which we call $\cA_i$ and $\cA_\s$ referring to all gauge fields collectively, including the graviton.
Then, since $\cA_i$ and $\cA_\s$ always appear together in the combination $\cA_i + \cA_\s$, 
the collinear and soft modes of $\cA$ are actually not separated.
(That is, even if \emph{we} write $\cA$ as $\cA_i + \cA_\s$, \emph{the theory} doesn't know it.) 

To solve this problem and accomplish true mode separation, 
we redefine every $\Phi_i$ schematically as
\beq
\Phi_i\too Y_{\! R_\Phi \!}[\cA_\s] \, \Phi_i
\,,\eql{BPS}
\eeq
where $R_\Phi$ denotes the gauge representations (including the spin) of the $\Phi_i$ and 
$\ds{Y_{\! R_\Phi \!}[\cA_\s]}$ is a functional of $\cA_\s$ (and also a function of $x$) such that the \emph{new} $\Phi_i$ after the redefinition is \emph{invariant} under $G_\s$.
Such field redefinition was first proposed and worked out for internal gauge symmetries in the context of the original QCD SCET~\cite{Bauer:2001yt}.
For gravity SCET, 
we will determine the form of $Y$ for diff$\times$Lorentz in \Sec{leading_soft_couplings_to_collinear_matter}.

With the field redefinition~\eq{BPS}, the effective gauge symmetry becomes truly factorized and mode separation completely achieved.
That is, rather than~\eq{eff_gauge_symm}, we now schematically have 
\beq
G \>\too\> 
G_\s \times G_1 \times G_2 \times \cdots \times G_N
\,,\eql{factorized_symm}
\eeq
which we call \emph{factorized effective gauge symmetry.}
Now, each field in the EFT is charged under only one of $G_\s$, $G_1$, \ldots, $G_N$.
It should be noted, however, that the implications of the ``$\ltimes$'' in~\eq{eff_gauge_symm} have not gone away completely. In particular, the structure~\eq{factorized_symm} inevitably double-counts soft modes and low-energy collinear modes.
This double-counting has to be systematically removed by \emph{zero-bin subtraction}~\cite{Manohar:2006nz}.
To the best of our knowledge, our theory is the first formulation of gravity SCET that is consistent with the factorized gauge symmetry (see~\Sec{comparison} for a comparison with the literature).

Finally, being true symmetries of nature rather than redundancies of our description, the \emph{global} part of $G$ (e.g, the global Poincar\'e group, the global $\SU(3)_\text{color}$) in the EFT is identical to that in the full theory.%
\footnote{To be clear, by gauge transformations we only refer to local transformations that vanish at infinity. So, global transformations are not a subgroup of gauge transformations.}
Namely, for each particle species, all of its $\Phi_i$ ($i = 1, \ldots, N$) and $\Phi_\s$ have the same, common, global charges as the corresponding full-theory $\Phi$. 
Only gauge symmetries are factorized in the EFT\@.

\subsection{The structure of the effective action}
\label{s.action_structure}
Although we adopt the ``position-space'' formulation of SCET~\cite{Beneke:2002ph, Beneke:2002ni} in our actual calculations (e.g., those in Appendices~\ref{a.Phi4}--\ref{a.ssff}), the language of Ref.~\cite{Freedman:2011kj} is convenient for expressing the SCET effective action $\cS_\text{eff}$ in a manner revealing mode separation and the symmetry structure~\eq{factorized_symm}:
\beq
\cS_\text{eff}[\Phi_1, \ldots, \Phi_N, \Phi_\s]
= \sum_{i=1}^N \cS_\text{full}[\Phi_i] + \cS_\text{full}[\Phi_\s] 
+ \cS_\text{hard}[\Phi_1, \ldots, \Phi_N, \Phi_\s]
\,,\eql{Seff_structure}
\eeq
Here, $\cS_\text{full}[\Phi_\s]$ is exactly the full-theory action $\cS_\text{full}[\Phi]$ with $\Phi$ replaced by $\Phi_\s$.
It is exactly the full-theory action because there is nothing ``soft'' about the soft sector in isolation, 
except that the running couplings in $\cS_\text{full}[\Phi_\s]$ should be evaluated at the scale $\mu \sim \la^2 Q$, corresponding to the only invariant energy scale of the soft sector, $p^2 \sim \la^4 Q^2$.
Similarly, there is nothing ``$n_i$-collinear'' about the $n_i$-collinear sector in isolation.
We could just boost the frame in the $\vec{n}_{\ps_i}$ direction by a rapidity of $\sim
\log\la$ so that the $n_i$-collinear scaling $\sim (1, \la^2, \la)_i$ would now scale isotropically as $\sim (\la, \la, \la)_i$.
Therefore, $\cS_\text{full}[\Phi_i]$ must be exactly the full-theory action with $\Phi$ replaced by $\Phi_i$,
except that the running couplings in $\cS_\text{full}[\Phi_i]$ should be evaluated at the scale $\mu \sim \la Q$, corresponding to the only invariant energy scale of the $n_i$-collinear sector, $p^2 \sim \la^2 Q^2$.
To be absolutely clear, the $\Phi_i$ in~\eq{Seff_structure} is the $\Phi_i$ after the field redefinition~\eq{BPS}, i.e., the one that is no longer charged under $G_\s$.
Finally, \emph{the hard interactions,} $\cS_\text{hard}[\Phi_1, \ldots, \Phi_N, \Phi_\s]$, consists of all possible terms containing more than one sector.

The factorized gauge symmetry structure~\eq{factorized_symm} is evident in the ``$\sum_i \cS_\text{full}[\Phi_i] + \cS_\text{full}[\Phi_\s]$'' part.
The entire non-triviality of SCET, therefore, is the implications of the factorized gauge symmetry and power counting for the structure of $\cS_\text{hard}$.

\subsection{Two types of renormalization}
The structure~\eq{Seff_structure} implies the following two categories of renormalization.
A diagram whose external lines all belong to the same sector renormalizes a vertex in $\cS_\text{full}[\Phi_\text{that sector}]$.
Such renormalization is already taken into account by evaluating the running couplings in the $\cS_\text{full}$ at the appropriate scale just discussed in \Sec{action_structure}.
Hence, no additional calculations are required in the EFT for such renormalizations. 

On the other hand, a diagram whose external lines belong to more than one sector renormalizes a vertex in $\cS_\text{hard}[\Phi_1, \ldots, \Phi_N, \Phi_\s]$.
The ``first'' instance of such renormalization is \emph{matching,} i.e., equating EFT and full-theory amplitudes at the hard scale $\mu \sim Q$
to determine the ``initial'' values of the coefficients of effective operators in $\cS_\text{hard}$.
Then, we have a series of incremental renormalizations from matching the EFT at scale $\ds{\mu-\dd\mu}$ onto the EFT at $\mu$, which gives us the renormalization group equations for those coefficients.

\subsection{Compatibility of power counting and gauge invariance}
\label{s.compatibility}
For any gauge field (including the graviton), the relative $\la$ dimensions between different components of the field can be completely fixed by requiring that power counting and gauge invariance be \emph{compatible} 
in the sense that the Ward identities should be satisfied order-by-order in $\la$ expansion.
For example, consider an $n_i$-collinear $\U(1)$ gauge field $A_\mu$ with
an $n_i$-collinear $\U(1)$ gauge invariance under $A_\mu \too A_\mu + \del_\mu \al$ with $\al$ only containing $n_i$-collinear Fourier modes.
This invariance leads to the Ward identity that an amplitude should vanish when the polarization vector $\vep_\mu(p)$ of a ``photon'' with an $n_i$-collinear momentum $p$ is replaced with the $p_\mu$. 
In order for such Ward identities to be satisfied order-by-order in $\la$,
the relative $\la$ dimensions between the components of an $\vep_\mu(p)$ must be the same as those between the components of the associated $p_\mu$.
That is, we must have $A^\mu(p) \sim \la^a p^\mu$,
i.e., $A \sim \la^a (1, \la^2, \la)_i$, with some $a$.
Moreover, gauge transformations must be able to gauge away the unphysical components of $A_\mu$, so we must have $\al \sim \la^a$ with the same $a$ so that $A_\mu \sim \del_\mu \al$.

The overall scaling parameter $a$ can then be fixed by requiring the kinetic term in the effective action to be $\cO(\la^0)$.
This is because our effective action is an expansion in powers of $\la$
and we are assuming that interactions are weak, i.e., that the kinetic terms dominate the action. 

Below, we will apply this line of reasoning to the collinear and soft graviton fields to determine how they scale with $\la$.

\subsection{Graviton fields and their scalings}
\label{s.gravitons}
We describe the spacetime geometry in terms of the vierbein $e^\mu_{~\nu}$, 
where the first index $\mu$ is a vector index for the Lorentz part of the diff$\times$Lorentz gauge group, 
and the second index $\nu$ is a 1-form index for the diff.  
By definition it satisfies $g_{\mu\nu} = \eta_{\rho\sg} e^\rho_{~\mu} e^\sg_{~\nu}$ at every spacetime point.
We write the inverse vierbein as $\bar{e}^\mu_{~\nu}$, 
where the first index $\mu$ is for the diff and the second $\nu$ is for the Lorentz. 
We have $\bar{e}^\mu_{~\rho} e^\rho_{~\nu} = \de^\mu_{\nu}$ and $\bar{g}^{\mu\nu} = \bar{e}^\mu_{~\rho} \bar{e}^\nu_{~\sg} \eta^{\rho\sg}$ by definition,
where $\bar{g}^{\mu\nu}$ is the inverse metric.%
\footnote{We use bars to distinguish the inverse vierbein and inverse metric from the vierbein and metric because of our convention stated in footnote~\ref{fn:metric}.}
We define the graviton field $\vphi^\mu_{~\nu}$ via
\beq
e^\mu_{~\nu} \equiv \de^\mu_\nu + \vphi^\mu_{~\nu}
\,.
\eeq
This then gives
\beq
\bar{e}^\mu_{~\nu} 
= \de^\mu_\nu - \vphi^\mu_{~\nu} + \cO(\vphi^2)
\,.
\eeq
We also define the metric fluctuation $h_{\mu\nu}$ via $g_{\mu\nu} \equiv \eta_{\mu\nu} + h_{\mu\nu}$.%
\footnote{Our graviton fields are not canonically normalized. 
To make them canonical, 
we must redefine the fields as $\vphi_{\mu\nu} \to \vphi_{\mu\nu} / \Mpl$ and $h_{\mu\nu} \to h_{\mu\nu} / \Mpl$,
where $\Mpl$ is the reduced Planck mass, $1 / \sqrt{8\pi G_\mathrm{N}}$.}
We then have $h_{\mu\nu} = \vphi_{\mu\nu} + \vphi_{\nu\mu} + \vphi_{\rho\mu} \vphi^{\rho}_{~\nu}$.
Most importantly, in the SCET,
mode separation tells us that the graviton field $\vphi_{\mu\nu}$ should be split into $N$ collinear graviton fields $\vphi^i_{\mu\nu}$ ($i = 1, \ldots, N$) and a soft graviton field $\vphi^\s_{\mu\nu}$.

Now let us find out how $\vphi^{i,\s}_{\mu\nu}$ scale with $\la$.
First, as we just discussed in \Sec{compatibility}, 
the relative $\la$ scalings between different components of $\vphi^{i,\s}_{\mu\nu}$ can be completely fixed by the compatibility of power counting and gauge symmetry.
Under an infinitesimal $n_i$-collinear or soft diff$\times$Lorentz gauge transformation, 
$\vphi^{i, \s}_{\mu\nu}$ transforms as
\beq
\vphi^{i, \s}_{\mu\nu} \too \vphi^{i, \s}_{\mu\nu} + \del_\nu \xi^{i, \s}_\mu + \om^{i, \s}_{\mu\nu} + \cdots
\eql{vphi_gauge-trans}
\eeq
where $\xi^{i, \s}_\mu(x)$ is the $n_i$-collinear or soft diff gauge transformation parameter,
$\omega^{i, \s}_{\mu\nu}(x)$ is an $n_i$-collinear or soft local Lorentz gauge transformation parameter,
and the ellipses represent terms containing both $\vphi^{i, \s}_{\mu\nu}$ itself and either one of the transformation parameters.
(We will justify why we can ignore the ellipses later.)
Then, the compatibility of power counting and local Lorentz gauge invariance tells us that we must have $\vphi^{i, \s}_{\mu\nu} \sim \la^a \om^{i, \s}_{\mu\nu}$ with some $a$.
Furthermore, we must be able to use this gauge transformation to gauge away the unphysical, anti-symmetric piece of $\vphi^{i, \s}_{\mu\nu}$ if we wish.
This requires $a=0$, and hence $\vphi^{i, \s}_{\mu\nu} \sim \om^{i, \s}_{\mu\nu}$.
Then, since $\om^{i, \s}_{\mu\nu}$ is anti-symmetric,
we also have $\vphi^{i, \s}_{\mu\nu} \sim \vphi^{i, \s}_{\nu\mu}$.

Similarly, the compatibility of power counting and diff gauge invariance tells us that $\vphi^{i, \s}_{\mu\nu} \sim \del_\nu \xi^{i, \s}_\mu$.
Here, since $\xi^i_\mu$ and $\xi^\s_\mu$ respectively contain only $n_i$-collinear and soft Fourier modes, 
the $\del_\nu$ acting on $\xi^i_\mu$ scales as $\del \sim p^i \sim (1, \la^2, \la)_i$, 
while the $\del_\nu$ on $\xi^\s_\mu$ as $\del \sim p^\s \sim (\la^2, \la^2, \la^2)$.
Therefore, we have $\vphi^{i, \s}_{\mu\nu} \sim p^{i, \s}_\nu \xi^{i, \s}_\mu$.
Then, recalling that $\vphi^{i, \s}_{\mu\nu} \sim \vphi^{i, \s}_{\nu\mu}$, 
we must have $p^{i, \s}_\nu \xi^{i, \s}_\mu \sim p^{i, \s}_\mu \xi^{i, \s}_\nu$,
which is possible only if $\xi^i_\mu \sim \la^b p^i_\mu$ and $\xi^\s_\mu \sim \la^c p^\s_\mu$ with some $b$ and $c$.
Thus, the compatibility of power counting and diff$\times$Lorentz gauge invariance completely fixes the relative scalings between the components of $\vphi^{i, \s}_{\mu\nu}$ and $\xi^{i, \s}_\mu$ as
\beq
\vphi^i_{\mu\nu} 
\sim \om^i_{\mu\nu} 
\sim \del_\mu \xi^i_\nu 
\sim \la^b p^i_\mu p^i_\nu
\,,\quad
\xi^i_\mu \sim \la^b p^i_\mu
\eql{temp-collinear-h-scaling}
\eeq
with $p^i \sim (1, \la^2, \la)_i$, and
\beq
\vphi^\s_{\mu\nu} 
\sim \om^\s_{\mu\nu} 
\sim \del_\mu \xi^\s_\nu 
\sim \la^c p^\s_\mu p^\s_\nu
\,,\quad
\xi^\s_\mu \sim \la^c p^\s_\mu
\eeq
with $p^\s \sim (\la^2, \la^2, \la^2)$.

To determine the overall scaling exponents $b$ and $c$, 
we need to look at the kinetic terms for $\vphi^{i,\s}_{\mu\nu}$ in the action and require them to be $\cO(\la^0)$ as we discussed in \Sec{compatibility}.
The kinetic terms for $\vphi^{i,\s}_{\mu\nu}$ in the effective action~\eq{Seff_structure} have the form
\beq
\cS_\text{full}[\vphi_{i, \s}] 
\propto 
\Mpl^2(\mu_{i, \s}) \int\! \dd^4x \, \del_\bullet \, \vphi^{i, \s}_{\bullet\bullet} \, \del^\bullet \, \vphi_{i, \s}^{\bullet\bullet}
\,,\eql{i-S-grav-kin}
\eeq
where different ways of contracting the spacetime indices $\bullet$ are summed over with some coefficients,
but such details are irrelevant for our discussion here.
The running reduced Planck mass $\Mpl(\mu_{i, \s})$ must be evaluated at the appropriate energy scale, $\mu_i \sim \la Q$ or $\mu_\s \sim \la^2 Q$, 
as we discussed in \Sec{action_structure}.%

Now, for the $n_i$-collinear graviton,
because of~\eq{temp-collinear-h-scaling} and the fact that  
contracting any two collinear momenta gives $\la^2$, 
the integrand of~\eq{i-S-grav-kin} scales as $\la^{2b} \la^6$. 
Furthermore, the $\dd^4x$ integration is dominated by the region of $x$ (after subtracting an overall spacetime translation) in which $\ds{p \dt x} = \ds{p^{\ps_i} x^{\ms_i} + p^{\ms_i} x^{\ps_i}+ p_{\pp_i} \dt x_{\pp_i}} \sim 1$ for $\ds{\e^{\I p \cdot x}}$. 
This implies that $x \sim (\la^{-2}, 1, \la^{-1})_i$, and therefore $\dd^4x \sim \la^{-4}$.
The above kinetic action thus scales as $\la^{2b} \la^2$.
Demanding this to be $\sim \la^0$ gives $b=-1$.
To summarize, we have
\beq
h^i_{\mu\nu} \sim \vphi^i_{\mu\nu} \sim \fr{p^i_\mu p^i_\nu}{\la} 
\eql{collinear-h-scaling}
\eeq
and
\beq
\xi^i_\mu \sim \fr{p^i_\mu}{\la}
\,,\quad
\om^i_{\mu\nu} \sim \fr{p^i_\mu p^i_\nu}{\la} 
\eql{collinear-xi-scaling}
\eeq
with $p^i \sim (1, \la^2, \la)_i$.

Let's repeat the same exercise for the soft graviton.
In this case, the integrand scales as $\la^{2c} \la^{12}$. 
The $x$ in~\eq{i-S-grav-kin} is now conjugate to a soft momentum, so it scales as $x \sim (\la^{-2}, \la^{-2},  \la^{-2})$, leading to $\dd^4 x \sim \la^{-8}$.
We then find $c=-2$, i.e., 
\beq
h^\s_{\mu\nu} \sim \vphi^\s_{\mu\nu} \sim \fr{p^\s_\mu p^\s_\nu}{\la^2} \sim \la^2
\eql{soft-h-scaling}
\eeq
and 
\beq
\xi^\s_\mu \sim \fr{p^\s_\mu}{\la^2} \sim \la^0
\,,\quad
\om^\s_{\mu\nu} \sim \fr{p^\s_\mu p^\s_\nu}{\la^2} \sim \la^2
\eql{soft-xi-scaling}
\eeq
with $p^\s \sim (\la^2, \la^2, \la^2)$.

In~\cite{Beneke:2012xa}, the scalings~\eq{collinear-h-scaling} and~\eq{soft-h-scaling} are determined by explicitly power-counting the propagator $\langle h_{\mu\nu} h_{\rho\sg} \rangle$ in a generic covariant gauge, painstakingly component-by-component in $\mu, \nu, \rho, \sg$ to obtain~\eq{collinear-h-scaling}.
Our derivation is much more efficient, thanks to the principle of compatibility of power counting and gauge invariance.
(See also comments in~\Sec{spin-1} for more on this last point.)

Finally, using the scaling relations obtained above, 
let's find out the $\la$ dimension of the ellipses in~\eq{vphi_gauge-trans}.
The ellipses consist of terms with one $\vphi_{\mu\nu}^{i,\s}$ and either one of $\del_\rho \xi_\sg^{i,\s}$ or $\om_{\rho\sg}^{i,\s}$ with two of the four indices being contracted.
So, the terms in the ellipses scale as $p_\mu p_\nu p_\rho p^\rho / \la^2 \sim p_\mu^{i} p_\nu^{i}$ in the collinear case,
and as $p_\mu p_\nu p_\rho p^\rho / \la^4 \sim p_\mu^{s} p_\nu^{s}$ in the soft case.
Thus, compared to the terms explicitly shown in~\eq{vphi_gauge-trans}, 
the ellipses are suppressed by $\la$ in the collinear case, and by $\la^2$ in the soft case. 
Therefore, even though the ellipses are also first order in the infinitesimal transformation parameters, we can still ignore them on the basis of $\la$ expansion.
In particular, we do not have to assume that $\vphi$ is also infinitesimal to justify our ignoring the ellipses in~\eq{vphi_gauge-trans}.
This will be important later when we derive some results that are valid to all orders in $1 / \Mpl$ (but at a fixed order in $\la$).

\subsection{Matter fields and their scalings}
\label{s.matter}
Let us quickly repeat the above exercise for matter (i.e., non-gravitational) fields for the sake of completeness.
The reader does not need this section to understand the rest of the paper and may skip to \Sec{nonlocal}.
For the cases of spin 0, $1/2$, and $1$, 
the results are well known in the literature
but our derivations of the results for the collinear spin $1/2$ and $1$ cases, which do not refer to any detailed forms of the lagrangians nor propagators, are not found in the existing literature to the best of our knowledge.
For the spin-$3/2$ case, our results also seem new.

\subsubsection{Spin 0}
For an $n_i$-collinear scalar field $\phi^i$, 
we have $\dd^4 x \sim \la^{-4}$ and $\del \del \sim \la^2$ in $\int \dd^4 x \, \del \phi^i \del \phi^i \sim \la^0$.
Thus, 
\beq
\phi^i \sim \la
\,.
\eeq
For a soft scalar field $\phi^\s$, we instead have $\dd^4 x \sim \la^{-8}$ and $\del \del \sim \la^4$, so
\beq
\phi^\s \sim \la^2
\,.
\eeq
%

\subsubsection{Spin $1/2$}
For a soft spinor field $\psi^\s$, demanding $\int \dd^4 x \, \psi^\s \del \psi^\s \sim \la^0$ with $\dd^4 x \sim \la^{-8}$ and $\del \sim \la^2$ tells us that 
\beq
\psi^\s \sim \la^3
\,.
\eeq
For an $n_i$-collinear spinor $\psi^i$, it is more complicated because the spatial non-isotropy of $n_i$-collinear momenta tells us that different spinor components of $\psi^i$ might scale differently.   
To skirt around this complication, let's first boost the frame in the $\vec{n}_{\ps_i}$ direction by a rapidity $\sim \log\la$ so that $n_i$-collinear momenta now scale isotropically as $\sim (\la, \la, \la)_i$ in the new frame.
In this frame, we simply have $\dd^4 x \sim \la^{-4}$ and $\del \sim \la$ in $\int \dd^4 x \, \psi^i \del \psi^i \sim \la^0$, implying that we have $\psi^i \sim \la^{3/2}$ for all components of $\psi^i$.
Now, if $\psi^i$ is a right-handed spinor, this boost multiplied the positive and negative helicity components of $\psi^i$ by a factor of $\la^{1/2}$ and $\la^{-1/2}$, respectively, when we got to the new frame.
So, the positive and negative helicity components of $\psi^i$ must have scaled as $\la$ and $\la^2$, respectively, in the original frame before the boost.
If $\psi^i$ is a left-handed spinor, 
the scalings of the positive and negative helicity components are simply reversed from the right-handed case, so the negative helicity component of $\psi^i$ scales as $\sim \la$ and the positive helicity as $\sim \la^2$.

To pick up the helicity components along the $\vec{n}_{\ps_i}$ direction from a spinor, we define the helicity projection operators:
\beq
P_i^+ \equiv \fr{\ga^{\ms_i} \ga^{\ps_i}}{2}
\,,\quad 
P_i^- \equiv \fr{\ga^{\ps_i} \ga^{\ms_i}}{2} 
\,,\eql{hel-projection}
\eeq
where our convention for the Dirac $\ga$ matrices is such that $\ga^\mu \ga^\nu + \ga^\nu \ga^\mu = 2\eta^{\mu\nu} \mathbbm{1}$.
Any spinor $\psi$ can then be decomposed as $\psi = P_i^+ \psi + P_i^- \psi$.
Since $P_i^\pm$ commutes with $\ga_5$, we can do these helicity projections separately for each chirality of $\psi$.
Then, we see that $P_i^+$ picks up the positive helicity component along the $\vec{n}_{\ps_i}$ direction from a right-handed spinor, and the negative helicity component from a left-handed spinor.
Whichever remaining helicity component is picked up by $P_i^-$.
Therefore,
the results we have found above can be summarized as
\beq
P_i^+ \psi^i \sim \la
\,,\quad
P_i^- \psi^i \sim \la^2
\,.
\eeq
Finally, with respect to an $n_j$-collinear direction with $j \neq i$, 
we have 
\beq
P_j^\pm \psi^i \sim \la
\,,\eql{fermion-cross-collinear}
\eeq
because different collinear sectors are well separated in direction
so both of $P_j^\pm$ pick up the bigger of the two components from $\psi^i$ (i.e., the one scaling as $\sim \la$).

Since the small components of spinors (i.e., those scaling as $\la^2$) have the wrong helicity to be on-shell, they are not propagating degrees of freedom.
It is therefore a common practice to integrate them out from the effective lagrangian.
However, for the purpose of keeping track of RPI, it is convenient to retain the small components in the lagrangian as auxiliary fields,
similarly to the method proposed by~\cite{Sundrum:1997ut} for maintaining manifest RPI in HQET\@.
(It is also morally similar to keeping the $F$ and $D$ components in a supersymmetric lagrangian for a better bookkeeping of supersymmetry.)
This is quite obvious for $\cS_\text{full}[\Phi_{i}]$, where there is nothing ``collinear'' about this action in isolation.
For $\cS_\text{hard}$, the small components can be ignored at the leading power but should be put back when we move on to the next-to-leading power to make RPI manifest and take into account constraints from RPI;
we will come back to this point later in \Sec{other_lambdas}.
Needless to say, the above comment also applies to the analogous small components of higher spin fields discussed below.

\subsubsection{Spin 1}
\label{s.spin-1}
Since we have assumed that our particles all have no or negligible mass, 
a spin-1 particle must be a gauge boson.
Then, as we already discussed in \Sec{compatibility}, 
the compatibility of power counting and gauge invariance tells us 
that $A_\mu^i \sim \la^a p^i_\mu$ for an $n_i$-collinear spin-1 gauge boson $A^i$ with a momentum $p^i \sim (1, \la^2, \la)_i$.
Similarly, we have $A_\mu^\s \sim \la^b p^\s_\mu$ for a soft spin-1 gauge boson $A^\s$ with $p^\s \sim (\la^2, \la^2, \la^2)$.
Then, the requirement $\int\! \dd^4 x \, \del A \del A \sim \la^0$ tells us that 
$a=b=0$, i.e.,
\beq
A^i_\mu \sim p^i_\mu 
\,,\quad
A^\s_\mu \sim p^\s_\mu
\,.\eql{spin-1-scaling}
\eeq

The above results are obtained in the literature by examining the explicit expression of the propagator $\langle A_\mu A_\nu \rangle$ in a generic Lorentz-covariant gauge, 
or by expanding the kinetic action explicitly in the light-cone coordinates without choosing the gauge and demanding that each term is $\sim \la^0$.
A conceptual advantage of our derivation based on the compatibility of power counting and gauge invariance is that it makes it clear why we want to work in a covariant gauge or without choosing the gauge initially,
because the compatibility would fail if a non-covariant gauge is imposed.
For example, let's choose a lightcone gauge by setting $A^{\ps_i} = 0$.
This completely kills diagrams that are proportional to the polarization $\ep^{\ps_i}(p)$. 
But these diagrams are the ones that would be the \emph{leading} diagrams in the Ward identity from $\ep^\mu(p) \to p^\mu$ as they would be proportional to $p^{\ps_i}$.
Thus, the $\la$ expansion of an amplitude does not agree with that of the corresponding Ward identity---hence incompatible in our language---in the light-cone gauge.
This does not necessarily mean that formulating a SCET in a non-covariant gauge is wrong,
but just that \emph{ensuring} or \emph{checking} gauge invariance of the theory would become a complicated problem (see~\cite{Idilbi:2010im, GarciaEchevarria:2011md} for a manifestation of the complications) because gauge invariance would relate terms of different orders in $\la$ in non-covariant gauges. 
This is why we have elevated the compatibility of power counting and gauge invariance to a guiding principle for constructing a SCET\@.

\subsubsection{Spin $3/2$}
Again, since our particles are all assumed to have no or negligible mass, 
a spin-$3/2$ field $\psi_\mu$ must be a gauge field, i.e., a gravitino. 
The gauge transformation (i.e., the supergravity transformation) has the form 
$\ds{\psi_\mu \to \psi_\mu + \DD_{\!\mu} \chi + \cdots}$, where the gauge transformation parameter $\chi$ is a spinor and the ellipses represent higher order terms analogous to the ellipses of~\eq{vphi_gauge-trans}.
Again, the compatibility of power counting and gauge invariance tells us that 
$\psi_\mu \sim p_\mu \chi$.
Then, for a soft gravitino $\psi^\s_\mu$, the $\la$ invariance of the kinetic action tells us that
\beq
\psi^\s_\mu \sim \la^3
\,,\quad
\chi^\s \sim \la
\,.\eql{soft-spin-3/2-scaling}
\eeq
For an $n_i$-collinear gravitino $\psi^i_\mu$, 
we use the same trick as we did for the spin-$1/2$ case and 
boost to an ``isotropic frame''.
In this frame, we have $\psi^i_\mu \sim \la^{3/2}$, and hence $\chi^i \sim \la^{1/2}$.
This means that
we must have had $P_i^+ \chi^i \sim \la^0$ and $P_i^- \chi^i \sim \la$ in the original frame before the boost. 
Therefore, from $\psi^i_\mu \sim p^i_\mu \chi^i$ in the original frame, we see that an $n_i$-collinear gravitino $\psi^i_\mu$ must scale as
\beqa{3}
P_i^+ \psi^i_\mu \sim p^i_\mu
\,,\quad
P_i^- \psi^i_\mu \sim \la p^i_\mu
\,,\eql{collinear-spin-3/2-scaling}
\eeqa
with $p^i \sim (1, \la^2, \la)_i$.
Finally, for $j \neq i$, we have 
\beq
P_j^\pm \psi^i_\mu \sim p^i_\mu
\eql{gravitino-cross-collinear}
\eeq
for the same reason mentioned for the similar relation for the spin $1/2$ case.

\subsection{Nonlocality in SCET}
\label{s.nonlocal}
Having discussed the symmetry structure~\eq{factorized_symm} 
and worked out the power counting rules for individual fields above, 
let us comment on a rather unusual feature of SCET, namely, its nonlocality.
In SCET, we actually have two types of nonlocality.
First, our power counting rules allow different fields to be located at different spacetime points in $\cS_\text{hard}$ as we will describe more precisely below in \Sec{nonlocality.1}.
Second, the mode separation implies the fields themselves as well as the (anti-)commutators among them are ``smeared''. This will be described in \Sec{nonlocality.2}.
Of course, despite these nonlocal building blocks, 
a SCET that is matched onto a local full theory is local.

\subsubsection{Nonlocality in hard interactions}
\label{s.nonlocality.1}
For a generic $n_i$-collinear field $\Phi_i(x)$,
we have $\del_{\ms_i} \Phi_i(x) = \del^{\ps_i} \Phi_i(x) \sim \la^0 \Phi_i(x)$.
This means that the $\la$ power counting does not allow the Taylor expansion of $\ds{\Phi_i(x^{\ps_i}, x^{\ms_i}\!+s, x_{\pp_i})}$ in powers of $s$ to be truncated at any finite order in $s$.
Therefore, different $n_i$-collinear fields with the same $i$ can have different $x^{\ms_i}$ coordinates in $\cS_\text{hard}$~\cite{Bauer:2000ew, Bauer:2000yr, Bauer:2001ct}.

On the other hand, since $\del_{\ps_i} \Phi_i \sim \la^2 \Phi_i$ and $\del_{\pp_i} \Phi_i \sim \la \Phi_i$, the Taylor expansions in the $x^{\ps_i}$ and $x_{\pp_i}$ coordinates can be (actually, must be, for manifest power counting) truncated at a finite order. So, all $n_i$-collinear fields in $\cS_\text{hard}$ must have the same $x^{\ps_i}$ and $x_{\pp_i}$ coordinates.%
\footnote{This would not hold true if we integrate out off-shell modes whose squared 4-momenta vanish in the $\la \to 0$ limit when we match the full theory to the EFT\@.
Such off-shell modes are not relevant for our target phase space
but become crucial if one considers highly energetic forward scattering processes (Regge limit)~\cite{Fleming:2014rea, Donoghue:2014mpa, Rothstein:2016bsq}.} 
This is in stark contrast to the situation in the more familiar Wilsonian EFTs in which all components of a derivative are equally suppressed by the cutoff $\La$, rendering the effective lagrangians completely local in all coordinates. 

As an example, imagine two $n_1$-collinear scalars $\phi_1$, $\chi_1$, and two $n_2$-collinear scalars $\phi_2$, $\chi_2$.
Then, $\cS_\text{hard}$ might look like this:
\beq
\int\!\dd^4x \, 
\dd s_{{}_1\!} \, \dd t_{{}_1\!} \, 
\dd s_{{}_2\!} \, \dd t_{{}_2\!} \, 
C(s_{{}_1}, t_{{}_1}, s_{{}_2}, t_{{}_2}) \,
\phi^\PD_{1\!} (x + s_{{}_1} n_{\ms_1\!}) \, 
\chi^\PD_{1\!} (x + t_{{}_1} n_{\ms_1\!}) \, 
\phi^\PD_{2\!} (x + s_{{}_2} n_{\ms_2\!}) \, 
\chi^\PD_{2\!} (x + t_{{}_2} n_{\ms_2\!})
\,,
\eeq
where the addition of $s_{{}_1} n_{\ms_1}^\mu$ to $x^\mu$ is displacing the $x^{\ms_1}$ to $\ds{x^{\ms_1\!} + s_{{}_1}}$, etc.
The ``Wilson coefficient'' $C(s_{{}_1}, t_{{}_1}, s_{{}_2}, t_{{}_2})$ may be determined by matching the EFT onto the full theory or related by symmetry to the Wilson coefficient of another operator.

The physical length scale of the nonlocality in the $x^{\ms_i}$ coordinate is $\cO(Q^{-1})$. 
Since our short-hand expression $p^{\ps_i} \sim \la^0$ actually means $p^{\ps_i} \sim Q$,
rapid oscillations of $\ds{\e^{\I p \cdot x}} \sim \ds{\e^{\I p^{\ps_i} x^{\ms_i}}}$ will damp the ``extra'' $\dd x^{\ms_i}$ integrations in $\cS_\text{hard}$ (such as $\dd s_1 \ldots \dd t_2$ in the above example)
once two fields get separated by a distance larger than $\sim Q^{-1}$ in the $x^{\ms_i}$ component.
This length scale itself is expected from the fact that we are integrating out off-shell modes whose virtuality is of $\cO(Q)$.
However, if our EFT were a Wilsonian EFT with a cutoff $\La \sim Q$, 
such nonlocality would not actually lead to nonlocal operators because all modes in the EFT would have wavelengths much longer than $\cO(Q^{-1})$ so would not be able to probe the nonlocality. 
In SCET, in contrast, there \emph{are} momentum modes with components of $\cO(Q)$,
so they can actually probe nonlocality of length scale of $\cO(Q^{-1})$.

\subsubsection{Nonlocality in field operators}
\label{s.nonlocality.2}
Due to mode separation, field operators in SCET are themselves nonlocal.
First, consider a soft scalar field $\phi_\s(x)$ and its canonical conjugate momentum $\pi_\s(x)$.
Since $\phi_\s$ and $\pi_\s$ only contain soft Fourier modes, 
their canonical commutation relation is given by
${}[\phi_\s(\vec{x}), \pi_\s(\vec{y})] = \I \de^3_\s (\vec{x} - \vec{y})$, 
where $\de^3_\s (\vec{x})$ is a 3-dimensional ``$\de$-function'' made only of soft Fourier modes rather than all Fourier modes.
Therefore, $\de^3_\s (\vec{x})$ is not exactly point-like at $\vec{x}=0$ but ``smeared''  over a length scale of $\cO(\la^{-2} Q^{-1})$, which is much \emph{larger} than $\cO(Q^{-1})$, i.e., the shortest length scale in the EFT\@.
The (anti-)commutation relations among collinear fields also have a similar, ``smeared'' kind of nonlocality in their respective $x^+$ and $x_\perp$ directions.
We will see in \Sec{leading_soft_couplings_to_collinear_matter} that the realization of the soft diff$\times$Lorentz symmetry through the field redefinition~\eq{BPS} actually ``exploits'' the smearing in the soft graviton field.

\section{Gravity SCET at the Leading Power (LP)}
\label{s.leading_order}
Let us now more explicitly construct gravity SCET at the Leading Power (LP\@).
When we talk about $\cS_\text{full}[\Phi_{i, \s}]$ in the effective action, 
LP literally refers to $\cO(\la^0)$ terms in $\cS_\text{full}$.
When we talk about $\cS_\text{hard}$, on the other hand, 
LP actually refers to the leading \emph{nontrivial} order in $\la$,
the precise meaning of which will become clear in \Sec{matching}.
Needless to say, when we say Next-to-Leading Power (NLP), that refers to one higher power of $\la$ compared to the LP\@.

\subsection{No LP graviton couplings within each collinear or soft sector}
\label{s.within}
We will first show the absence of $\cO(\la^0)$ gravitational interactions in $\cS_\text{full}[\Phi_{i, \s}]$ in the effective action~\eq{Seff_structure},
based only on mode separation, power counting, and the factorized gauge symmetry~\eq{factorized_symm}.

\subsubsection{No LP graviton couplings within each collinear sector}
\label{s.no_collinear_gravity_at_leading_order}
The mode separation~\eq{Seff_structure} and $\la$ power counting tell us that the $n_i$-collinear graviton field has no $\cO(\la^0)$ interactions in $\cS_\text{full}[\Phi_i]$.
As we already discussed, $\cS_\text{full}[\Phi_i]$ is exactly the full theory action at the energy scale $\sim \la Q$,
containing all vertices that only join $n_i$-collinear lines. 
In particular, there is nothing ``$n_i$-collinear'' about $\cS_\text{full}[\Phi_i]$ in isolation 
because we can boost in the $\vec{n}_{\ps_i}$ direction by a rapidity of $\sim \log\la$ to a frame in which what used to scale as $\sim (1, \la^2, \la)_i$ in the original frame now scales as $\sim (\la, \la, \la)_i$.
(This is a \emph{global} Lorentz boost so the graviton field also transforms covariantly just like everyone else.) 
If we do dimensional analysis in such a frame, 
each mass dimension of the fields and derivatives in $\cS_\text{full}[\Phi_i]$ simply counts as $\la Q$, because this is the only dynamical scale in $\cS_\text{full}[\Phi_i]$.
Then, since every gravitational interaction comes with a positive power of $1 / \Mpl$, 
it comes with a positive power of $\la Q$, thereby vanishing as $\la \to 0$.
We thus clearly see that $\cS_\text{full}[\Phi_i]$ has no LP $n_i$-collinear graviton couplings to $n_i$-collinear particles, including $n_i$-collinear gravitons themselves. 
Diagrammatically, an $n_i$-collinear graviton line can never be attached to any $n_i$-collinear line with the same $i$ at the LP\@.

\subsubsection{No soft graviton couplings within the soft sector at the LP}
\label{s.soft_selfcouplings_order_one}
Similarly, 
the mode separation and $\la$ power counting tell us that the soft graviton has no interaction terms of $\cO(\la^0)$ in $\cS_\text{full}[\Phi_\s]$.
From the same argument as above (but without the need for a boost), 
we see that every gravitational interaction scales as a positive power of $\la^2 Q$, thereby vanishing as $\la \to 0$.
Therefore, there are no $\cO(\la^0)$ nor $\cO(\la)$ soft graviton couplings to soft particles in $\cS_\text{full}[\Phi_\s]$,
including soft gravitons themselves.
Diagrammatically, a soft graviton line can never be attached to any soft line both at the LP and NLP\@.

\subsection{Building blocks of hard interactions}
\label{s.hard}
Let us now talk about $\cS_\text{hard}[\Phi_1, \ldots, \Phi_N, \Phi_\s]$ in the effective action~\eq{Seff_structure}.

\subsubsection{Matching}
\label{s.matching}
The very first step for constructing $\cS_\text{hard}[\Phi_1, \ldots, \Phi_N, \Phi_\s]$ is matching.
To perform matching, turn off all the interactions in the EFT that are \emph{forced upon us} by the factorized gauge symmetry~\eq{factorized_symm}.
Let's call this limit \emph{purely hard.}
The purely hard limit still leaves the \emph{global} part of $G$---e.g., the global Lorentz invariance---completely intact in the EFT\@. 
The limit also keeps all interactions that are not \emph{required} by~\eq{factorized_symm}.%
\footnote{For example, the limit turns all covariant derivatives into ordinary partial derivatives $\del$, 
but leaves the leading terms of gauge-invariant objects, e.g.~the 2-graviton terms in $\phi^2 R_{\mu\nu} R^{\mu\nu}$, completely untouched as their coefficients are not fixed by gauge invariance.
On the other hand, the terms with 3 or more gravitons in $\phi^2 R_{\mu\nu} R^{\mu\nu}$ would be discarded in the purely hard limit.}
The purely hard limit is not applied to the full theory.

Upon matching, purely hard amplitudes from the EFT are set equal to the corresponding amplitudes from the full theory.
This determines $\cS_\text{pure}[\Phi_1, \ldots, \Phi_N]$, that is, 
the purely hard part of $\cS_\text{hard}[\Phi_1, \ldots, \Phi_N, \Phi_\s]$.
Note that there is no $\Phi_\s$ in $\cS_\text{pure}[\Phi_1, \ldots, \Phi_N]$
because, as stated in \Sec{target}, a soft particle is there only when it is ``required by nature''---which in the present context means gauge invariance---so the purely hard limit excludes soft particles from consideration.
The warning we mentioned at the beginning of \Sec{leading_order} regarding the meaning of ``LP'' for $\cS_\text{hard}$ can now be stated more clearly: When we talk about $\cS_\text{hard}$, LP refers to the lowest $\la$ dimension in $\cS_\text{pure}$.
 
Matching in the purely hard limit can be done at any desired order in the number of loops and coupling constants of the full theory, such as $1 / \Mpl(Q)$. 
In the following analyses, we will see that once $\cS_\text{pure}$ is given, 
$\cS_\text{hard}$ can be uniquely constructed at the LP, thanks to the factorized gauge symmetry~\eq{factorized_symm}.

\subsubsection{Collinear diff$\times$Lorentz invariant objects at the LP}
\label{s.collinear_invariance}
We first discuss collinear graviton interactions in $\cS_\text{hard}[\Phi_1, \ldots, \Phi_N, \Phi_\s]$.
Let $\cO_i$ be the product of all $n_i$-collinear objects in any one of the terms in $\cS_\text{hard}[\Phi_1, \ldots, \Phi_N, \Phi_\s]$, 
where $\cO_i$ may be a single $n_i$-collinear field or the product of $n_i$-collinear fields with or without derivatives, $\ga$-matrices, etc.
A key point is that the factorized effective gauge symmetry~\eq{factorized_symm} 
requires $\cO_i$ to be not only invariant under all $G_j$ with $j \neq i$ but also under $G_i$ itself, because nothing else other than $\cO_i$ itself is charged under $G_i$ within the same term of $\cS_\text{hard}$.
This seemingly trivial requirement turns out to be extremely powerful for constraining possible hard interactions, going much further than the constraints from the global part of $G$ common to all sectors.

Consider an arbitrary $n_i$-collinear local operator $\Phi_i^{(r)\!}(x)$ that is a scalar under the $n_i$-collinear diff group and transforms covariantly as a representation $r$ under the $n_i$-collinear local Lorentz group. 
We make it a diff scalar even for an integer-spin $r$ by appropriately multiplying it by vierbeins.
If derivatives are acting on an $n_i$-collinear field, 
we include all of them inside $\Phi_i^{(r)}$, again with appropriate vierbeins multiplying them to make them diff scalars.
In this way, we can treat the bosonic and fermionic cases and derivatives simultaneously.

Now, because of the factorized gauge symmetry~\eq{factorized_symm}, $\Phi_i^{(r)}$ is already invariant under all $n_j$-collinear diff$\times$Lorentz gauge group with $j \neq i$ to all orders in $\la$.
Remarkably, $\Phi_i^{(r)}$ is also invariant at the LP under the $n_i$-collinear diff$\times$Lorentz gauge group of the same $i$.
If $\Phi_i^{(r)}$ is literally just the field operator of an $n_i$-collinear particle, 
this invariance is trivially implied by 
the absence of LP $n_i$-collinear graviton couplings to $n_i$-collinear particles discussed in \Sec{no_collinear_gravity_at_leading_order}.
However, as we said above, $\Phi_i^{(r)}$ may contain derivatives and vierbeins in addition to the ``elementary'' field operator. 
Therefore, we must show the invariance of $\Phi_i^{(r)}$ without referring to what it is made of. 
So, let's directly examine an infinitesimal $n_i$-collinear diff$\times$Lorentz gauge transformation on $\Phi_i^{(r)}$:
\beq
\Phi_i^{(r)} \too \Phi_i^{(r)} + \xi_i^{\mu} \del_\mu \Phi_i^{(r)} + \fr12 \om^i_{\mu\nu} \Sg_r^{\mu\nu} \Phi_i^{(r)}
\,,\eql{collinear-matter-collinear-transform}
\eeq
where $\Sg_r^{\mu\nu}$ are the Lorentz generators for the representation $r$,
satisfying the algebra
\beq
{}[ \Sg_r^{\al\be}, \Sg_r^{\ga\de} ] 
= \eta^{\al\de} \Sg_r^{\be\ga} + \eta^{\be\ga} \Sg_r^{\al\de}  
- \eta^{\al\ga} \Sg_r^{\be\de} - \eta^{\be\de} \Sg_r^{\al\ga}
\eql{lorentzalgebra}
\eeq
with $(\eta^{\al\be}) = \diag(1,-1,-1,-1)$ and without a factor of $\I$ on the right-hand side.
According to~\eq{collinear-xi-scaling}, the term with $\xi_i^\mu$ in~\eq{collinear-matter-collinear-transform} scales as $(p_i^\mu / \la) \, p^i_\mu \Phi_i^{(r)} \sim \la^2 / \la \Phi_i^{(r)} \sim \la \Phi_i^{(r)}$,
so it should be discarded at the LP\@.
It turns out that the $\om^i_{\mu\nu}$ term also scales as $\sim \la \Phi_i^{(r)}$ and hence should be discarded.
To see this, let's boost in the $\vec{n}_{\ps_i}$ direction by a rapidity of $\sim \log\la$ so that $n_i$-collinear momenta now scale as $\sim (\la, \la, \la)_i$. 
Then, the scaling law~\eq{collinear-xi-scaling} in the boosted frame becomes $\om^i_{\mu\nu} \sim \la \la / \la \sim \la$.
Since the expressions $\Phi_i^{(r)}$ and $\om^i_{\mu\nu} \Sg_r^{\mu\nu} \Phi_i^{(r)}$ are both Lorentz covariant, 
the fact that the latter is suppressed by $\la$ compared to the former must hold true in the original frame as well.
We thus conclude that $\Phi_i^{(r)}$ does not transform at all at the LP under the $n_i$-collinear diff$\times$Lorentz gauge group.
Therefore, $\Phi_i^{(r)}$ is already completely gauge invariant at the LP under all collinear diff$\times$Lorentz groups, 
and hence it may by itself form $\cO_i$.%
\footnote{Note that $\Phi_i^{(r)}$ still \emph{does} transform as the representation $r$ under the \emph{global} Lorentz group.
But the global Lorentz invariance can be taken care of by contracting, for example, a vector index from sector $i$ with one from sector $j$, so it is not useful for constraining the structure of $\cO_i$ in isolation.
The requirement of invariance of $\cO_i$ only refers to the factorized gauge symmetry, not to the global symmetry.}

It is possible that a purely hard process may have an $n_i$-collinear graviton.
However, the $n_i$-collinear graviton field $\vphi^i_{\mu\nu}$ cannot be a $\Phi_i^{(r)}$  
because, being a gauge field, $\vphi^i_{\mu\nu}$ does not transform \emph{covariantly} under the local Lorentz group.
In fact, unlike $\Phi_i^{(r)}$, $\vphi^i_{\mu\nu}$ \emph{does} transform at the LP as one can directly see in~\eq{vphi_gauge-trans} combined with~\eq{collinear-h-scaling} and~\eq{collinear-xi-scaling}.
(One should also recall that those scaling laws come partly from the requirement that gauge fields in general \emph{should} transform at the LP so that their unphysical polarizations can be gauged away at the LP\@.)
Therefore, a ``bare'' $\vphi^i_{\mu\nu}$ is not factorized gauge invariant and hence cannot appear in $\cS_\text{hard}$.
But we can find a covariant object $\Phi_i^{(r)}$ that \emph{contains} a $\vphi^i_{\mu\nu}$.
There are three possible covariant objects that can create or annihilate one graviton with the fewest possible derivatives.
The simplest possibility is the $n_i$-collinear Ricci scalar $R^i$ made of $h^i_{\mu\nu}$.
Since 1-graviton terms in $R^i$ form a scalar built out of two derivatives and a collinear graviton field, they scale as $p_\mu p_\nu p^\mu p^\nu / \la$ with an $n_i$-collinear momentum $p$, so we have
\beq
R^i \sim \la^2 \la^2 / \la = \la^3
\,.\eql{Ri}
\eeq
The next simplest object is the $n_i$-collinear Ricci tensor $R^i_{\mu\nu}$.
The 1-graviton terms in it scale as $p_\mu p_\nu p_\rho p^\rho / \la$ with an $n_i$-collinear momentum $p$, so we have
\beq
R^i_{\mu\nu} \sim \la p_\mu p_\nu
\,,\eql{Rimunu}
\eeq
of which the largest component is $R^i_{\ms_i \ms_i} \sim \la$ because $p_{\ms_i} = p^{\ps_i} \sim \la^0$.
Similarly, the 1-graviton terms in the $n_i$-collinear Riemann tensor $R^i_{\mu\nu\rho\sg}$ scale as
\beq
R^i_{\mu\nu\rho\sg} \sim \fr{p_{[\mu} p_{\nu]} p_{[\rho} p_{\sg]}}{\la}
\,.\eql{Rimunurhosigma}
\eeq
Beware of the antisymmetry in each of the $\mu$-$\nu$ and $\rho$-$\sg$ pairs.
So, the largest component is given by $R^i_{\ms_i \pp_i \ms_i \pp_i} \sim \la \la / \la = \la$.
In all the three cases above, 2-graviton terms are further suppressed by an extra $\ds{p \dt p} / \la \sim \la$ (and, thus, every extra collinear graviton field costs an additional $\la$).
Any of these three objects, \eq{Ri}, \eq{Rimunu}, or \eq{Rimunurhosigma}, can be a $\Phi_i^{(r)}$ and hence can form an $\cO_i$ by itself.
Other two-derivative covariant objects, such as the conformal or Weyl tensor $C^i_{\mu\nu\rho\sg}$, are linear combinations of these three objects.
These are all local operators, but since nonlocality in the $x^{\ms_i}$ direction is allowed in SCET, we can also construct nonlocal invariant operators by integrating~\eq{Ri}, \eq{Rimunu}, or \eq{Rimunurhosigma} in the $x^{\ms_i}$ direction,
thereby providing us with invariant objects with fewer derivatives than two.
But such constructions are already implicit in the general nonlocal structure of SCET described in \Sec{nonlocality.1} and hence not new given~\eq{Ri}, \eq{Rimunu}, \eq{Rimunurhosigma}.
One may also wonder if there are other nonlocal invariant objects built out of $\vphi^i_{\mu\nu}$. There \emph{are}---they are called Wilson lines---but it turns out that they are nontrivial only if we proceed to the NLP, so they will be discussed in \Sec{order-lambda}.  

The fact that $\Phi_i^{(r)}$ is invariant under all collinear diff$\times$Lorentz gauge groups at the LP in particular means that there are no graviton couplings from the covariant derivatives that may be ``hiding'' in $\Phi_i^{(r)}$.
For example, suppose $\Phi_i^{(r)} = \ds{\DD_{\!\mu} \Phi_i^{(r')}}$.
By our argument above,  
$\Phi_i^{(r')}$ is invariant under all diff$\times$Lorentz groups at the LP,
provided that it is a covariant object itself.%
\footnote{So, $\Phi_i^{(r')}$ cannot be a ``bare'' graviton field but can be \eq{Ri}, \eq{Rimunu}, or \eq{Rimunurhosigma}.}
The field redefinition~\eq{BPS} makes it also invariant under the soft diff$\times$Lorentz group.
Therefore, as far as gravity is concerned, the $\DD_{\!\mu}$ just becomes a $\del_\mu$ at the LP\@.

We should also recall that $\Phi_i^{(r)}$ may contain vierbeins.
However, again, there are no LP collinear graviton couplings from those vierbeins.
To see this, note that the vierbeins are placed to convert diff vector indices to Lorentz vector indices. 
So, imagine an expression of the ``inside'' of $\Phi_i^{(r)}$ in terms of Lorentz indices only. 
In this expression, there are no \emph{explicit} vierbeins. 
The Lorentz vector indices carried by constant objects like the $\gamma$ matrices have no vierbeins, and hence no gravitons, in them.
On the other hand, the Lorentz vector indices from ``naturally diff'' indices---i.e., those of derivatives and of fields with spin $\geq 1$---actually contain vierbeins and, hence, gravitons.
However, the couplings of these gravitons are at most NLP\@.
For example, consider $\del_\mu$ with a Lorentz index $\mu$, which contains a graviton coupling of the form $\phi_\mu^{i \, \nu} \del_\nu$.
From the scaling law~\eq{collinear-h-scaling}, we have
$\vphi_\mu^{i\,\nu} \del_\nu \sim p_\mu p^\nu p_\nu / \la \sim p_\mu \la$,
where $p$ is an $n_i$-collinear momentum.
In the worst case, this $p_\mu$ would be contracted with a momentum in another collinear sector and thus counts as LP due to the cross-collinear scaling~\eq{cross-collinear}.
So, $\phi_\mu^{i \, \nu} \del_\nu$ is suppressed at least by $\la$. 
If the $\del_\nu$ above is replaced by a spin-1 gauge field, the conclusion is unchanged as spin-1 fields scale just like momenta (see~\eq{spin-1-scaling}).
If the $\del_\nu$ is replaced by a spin-$3/2$ field, it is again the same conclusion (see~\eq{collinear-spin-3/2-scaling}).
Finally, if it is replaced by one of the indices of~\eq{Rimunu} or \eq{Rimunurhosigma}, it is again suppressed by at least $\la$. 
Therefore, at the LP, all vierbeins inside $\Phi_i^{(r)}$ should be replaced by Kronecker deltas.
Combining this with the observation of the preceding paragraph, 
we conclude that there are no LP couplings of $n_i$-collinear gravitons ``hiding'' inside $\Phi_i^{(r)}$.

This might be a good place to discuss the implications of the \emph{global} Lorentz invariance common to all sectors.
For example, 
the way the indices of $R^i_{\mu\nu}$ and $R^i_{\mu\nu\rho\sigma}$ appear in $\cS_\text{hard}$ must obey the global Lorentz invariance.
So, for example, $R^i_{\ms_i \ms_i}$ must be part of a globally Lorentz invariant object such as $(n_{\ms_i})^\mu (n_{\ms_i})^\nu R^i_{\mu\nu}$ and $R^i_{\mu\nu} \times (\text{an $n_j$-collinear operator)}^{\mu\nu}$, where the cross-collinear scaling~\eq{cross-collinear} should be applied to the latter to get $R^i_{\ms_i \ms_i}$.
Needless to say, such considerations must be applied to all vector/spinor indices in $\cS_\text{hard}$, not just those of the curvature tensors. 

To summarize, we first start with the purely hard limit, $\cS_\text{pure}[\Phi_1, \ldots, \Phi_N]$, matched at the leading nontrivial order in $\la$.
We then turn on all collinear diff$\times$Lorentz gauge groups of the factorized effective gauge symmetry~\eq{factorized_symm} (but $G_\s$ still turned off).
At the LP, this does not change anything, i.e., 
we just have $\cS_\text{hard}[\Phi_1, \ldots, \Phi_N] = \cS_\text{pure}[\Phi_1, \ldots, \Phi_N]$ at the LP\@.
There are no collinear graviton couplings here.

\subsubsection{Soft graviton couplings in $\cS_\text{hard}$ at the LP}
\label{s.leading_soft_couplings_to_collinear_matter}
Now, let's finally turn on $G_\s$.
As we discussed in \Sec{fact_symm}, 
this amounts to performing the field redefinition~\eq{BPS} for each $\Phi_i$ in $\cS_\text{hard}[\Phi_1, \ldots, \Phi_N]$ obtained above (that is, just $\cS_\text{pure}[\Phi_1, \ldots, \Phi_N]$ itself). 
So, the goal of this section is to determine the form of the $Y$ functional in~\eq{BPS} at the LP\@. 
Note that, by the definition of the purely hard limit introduced in \Sec{matching}, the soft graviton couplings we get in going from $\cS_\text{pure}$ to $\cS_\text{hard}$ are just those required by the soft gauge invariance.
We will see that $Y$ is completely determined by the soft gauge invariance alone at the LP\@.

Consider again an arbitrary $n_i$-collinear, Lorentz-covariant, diff-scalar operator $\Phi_i^{(r)}$ of representation $r$ of the $n_i$-collinear local Lorentz group, 
as we did in \Sec{collinear_invariance}.
First,
let $\Phi_i^{(r)}$ refer to the ``original'' collinear field before the field redefinition~\eq{BPS},
i.e., the one that is still charged under $G_\s$. 
So, analogous to~\eq{collinear-matter-collinear-transform},
we have an infinitesimal soft diff$\times$Lorentz gauge transformation: 
\beq
\Phi_i^{(r)} \too \Phi_i^{(r)} + \xi_\s^{\mu} \del_\mu \Phi_i^{(r)} + \fr12 \om^\s_{\mu\nu} \Sigma_r^{\mu\nu} \Phi_i^{(r)}
\,.\eql{collinear-matter-soft-transform}
\eeq
Here, unlike in~\eq{collinear-matter-collinear-transform}, 
$\Phi_i^{(r)}$ does transform at the LP\@. 
In particular, 
the $\xi_\s^{\ms_i} \del_{\ms_i}$ piece inside the $\xi_\s^{\mu} \del_\mu$ term above is LP, 
because we have $\xi_\s^{\mu} \sim \la^0$ from~\eq{soft-xi-scaling} and the $\del_\mu$ above scales as an $n_i$-collinear momentum so $\del_{\ms_i} = \del^{\ps_i} \sim \la^0$.
The remaining terms in $\xi_\s^{\mu} \del_\mu$ are NLP or higher.
All components of $\om^\s_{\mu\nu}$ scale as $\sim\la^2$ from~\eq{soft-xi-scaling}.
Here, without any calculations, we immediately see 
two well-known pieces of physics.
First, soft graviton couplings at the LP are all spin independent, i.e., independent of $r$, 
because the action of $\xi_\s^{\mu} \del_{\mu}$ on $\Phi_i^{(r)}$ does not depend on $r$. 
This leads to the second point that spin dependence must be a Next-to-Next-to-Leading Power (NNLP) effect as it is only associated with the $\om^\s_{\mu\nu}$ term in~\eq{collinear-matter-soft-transform}.

To determine the $Y$ functional in~\eq{BPS} at the LP, 
let's focus on the LP piece of~\eq{collinear-matter-soft-transform}:   
\beq
\Phi_i \too \Phi_i + \xi^\s_{\ps_i} \del_{\ms_i} \Phi_i
\,,\eql{Phi_i-leading-soft-transform}
\eeq
where we have dropped the ${}^{(r)}$ because spin does not matter at this order.
Therefore, in order for the field redefinition~\eq{BPS} to take care of the transformation \eq{Phi_i-leading-soft-transform}, 
all we care about is the $\xi^\s_{\ps_i}$ piece of the soft diff$\times$Lorentz.
That is, at the LP, we want to have
\beq
Y_i[\vphi^\s] \too Y_i[\vphi^\s] + \xi^\s_{\ps_i} Y_i[\vphi^\s] \, \del_{\ms_i} 
\eql{Y_i-trans}
\eeq
with the soft graviton field $\vphi^\s$.
To find such an object, let us return to the soft diff$\times$Lorentz gauge transformation~\eq{vphi_gauge-trans} for $\vphi^\s_{\mu\nu}$.
There, one sees that, in order to have any chance of getting a $\xi^\s_{\ps_i}$ to match~\eq{Y_i-trans}, the $\mu$ of $\vphi^\s_{\mu\nu}$ in~\eq{vphi_gauge-trans} must be $+_i$. 
Next, we do not want to have any $\om^\s_{\mu\nu}$ from~\eq{vphi_gauge-trans},
because there is no $\om^\s_{\mu\nu}$ in~\eq{Y_i-trans}.
Since there is no power-counting reason to throw away $\om^\s_{\mu\nu}$ in~\eq{vphi_gauge-trans}, the only way not to have it is by having $\nu = \mu$. 
Hence, the only relevant component of~\eq{vphi_gauge-trans} is
\beq
\vphi^\s_{\ps_i \ps_i} \too \vphi^\s_{\ps_i \ps_i} + \del_{\ps_i} \xi^\s_{\ps_i}
\,.\eql{plusplus}
\eeq
Since $\xi^\s_{\ps_i}$ in~\eq{Y_i-trans} is not differentiated by a $\del_{\ps_i}$,
we must integrate $\vphi^\s_{\ps_i \ps_i}$ by $\dd x^{\ps_i}$ to remove the $\del_{\ps_i}$ in~\eq{plusplus}.
Therefore, we see that the structure
\beq
\exp\lt[ \pm \int_{x_1}^{x_2} \!\dd x^{\ps_i} \, \vphi^\s_{\ps_i \ps_i \!} (x^{\ps_i}) \, \del_{\ms_i}\rt]\!
\eql{Y_rough_form} 
\eeq
has exactly the right transformation property to be $Y_i[\vphi^\s]$, 
provided that we choose the sign and limits of integration appropriately to exactly match~\eq{Y_i-trans}.
(The coordinates $x^{\ms_i}$ and $x_{\pp_i}$ are implicit in \eq{Y_rough_form}.)
An obvious choice is to place $x_2$ at the point $x$ where the $\Phi_i(x)$ that $Y_i$ is acting on is located.
Then, we must choose the $+$ sign in~\eq{Y_rough_form} and $x_1$ must be placed at a past infinity, $x_1^{\ps_i} \to {-\infty}$, so that we do not get any contribution to~\eq{Y_i-trans} from the $x_1$ end of the integral. 
Namely,
\beq
\exp\lt[ \,\int_{-\infty}^0\!\! \dd s \, \vphi^\s_{\ps_i \ps_i \!} (x + s n_{\ps_i}) \, \del_{\ms_i} \rt],
\eql{past} 
\eeq
where the addition of $s n_{\ps_i}^\mu$ to $x^\mu$ is shifting $x^{\ps_i}$ by $s$. 
Alternatively, we can place $x_1$ at $x$, choose the $-$ sign in~\eq{Y_rough_form} and send $x_2$ to a future infinity, $x_2^{\ps_i} \to {\infty}$.
That is, 
\beq
\exp\lt[ -\!\int_0^\infty\!\! \dd s \, \vphi^\s_{\ps_i \ps_i \!} (x + s n_{\ps_i}) \, \del_{\ms_i} \rt].
\eql{future} 
\eeq
Which solution should we use? 
In field theory, when we integrate over energy $p^0$ from $-\infty$ to $\infty$, we must rotate the contour infinitesimally counterclockwise in the complex $p^0$ plane to ensure that it is always the positive energy wave that propagates to the future. 
Now, suppose we have a soft graviton line with its energy going into a vertex from $Y_i$.
This graviton is annihilated at the vertex by the positive frequency part of $\ds{\vphi^\s_{\ps_i \ps_i \!} (x + s n_{\ps_i})}$, which goes as $\ds{\exp[-\I p^{\ps_i} s]}$ in terms of the graviton's momentum $p^{\ps_i} > 0$ and integration variable $s$.
Then, in order for the integration over $s$ to converge after $p^{\ps_i}$ is replaced by $\ds{p^{\ps_i} + \I\ep}$ with an infinitesimal positive $\ep$, we must send $s$ to $-\infty$.
If the soft graviton's energy is going out of the vertex from $Y_i$, 
we must send $s$ to $+\infty$ instead.
Thus, combining both cases, we arrive at the expression
\beq
Y_i(x) = \exp\lt[ -\!\int'\!\! \dd s \, \vphi^\s_{\ps_i \ps_i \!} (x + s n_{\ps_i}) \,\del_{\ms_i} \rt],
\eql{softWilson}
\eeq
where for any quantum field $f(s) = f_+(s) + f_-(s)$ with $f_+(s)$ and $f_-(s)$ being the positive and negative frequency parts, respectively, and $s$ being a shift in one of the coordinates,  
we define $\int'\!\dd s$ as
\beq
\int'\!\! \dd s \, f(s) \equiv 
-\!\int_{-\infty}^0\!\! \dd s \, f_+(s) 
+\!\int_0^\infty\!\! \dd s \, f_-(s)
\,.\eql{Is}
\eeq
As we argued above, this splitting is necessary so that we can Wick-rotate the energy components of all momentum integration variables consistently in the counterclockwise direction in the complex energy plane.%
\footnote{\label{fn:splitting}Nevertheless, in practice we may ignore the splitting and just adopt either one of~\eq{past} and~\eq{future} for $Y_i$, because even if one forgets about $\I\ep$ everyone knows which way to rotate the contour when it comes a time to do a Wick rotation.}

Let's check explicitly that the soft graviton couplings from $Y_i$ to $n_i$-collinear particles are indeed LP\@.
First, in the above expressions for $Y_i$, the derivative~$\del_{\ms_i}$ should not act on $\ds{\vphi^\s_{\ps_i \ps_i \!}(x + s n_{\ps_i})}$ because, if it did, it would give $\sim \la^2$ due to the scaling law~\eq{soft-h-scaling}.
Rather, it should only act on the $n_i$-collinear field $\Phi_i$ that $Y_i$ acts on.
Then, $\del_{\ms_i}$ in~\eq{softWilson} counts as $\sim \la^0$ because $\del_{\ms_i} \Phi_i \sim \la^0 \Phi_i$.
Next,
to understand how $\dd s$ in~\eq{softWilson} scales,
it is important to return to the field redefinition~\eq{BPS} to recognize that the $x$ of $Y_i(x)$ is the $x$ of the $n_i$-collinear field $\Phi_i(x)$, not an $x$ of a soft field.
So, we have $(\dd x^{\ps_i}, \dd x^\ms_i, \dd x_{\pp_i}) \sim (1 / p^{\ms_i}, 1 / p^{\ps_i}, 1 / p_{\pp_i}) \sim (\la^{-2}, 1, \la^{-1})_i$ 
and hence $\dd s \sim \dd x^{\ps_i} \sim \la^{-2}$.
Finally, we have $\vphi^\s_{\ps_i \ps_i} \sim \la^2$ from~\eq{soft-h-scaling}. 
We thus see that the exponent in~\eq{softWilson} is indeed LP\@.
Combining this with the above symmetry argument that led to the expression~\eq{Y_rough_form},
we conclude that $Y_i$ gives us all LP couplings of soft gravitons in $\cS_\text{hard}$ that are correct to all orders in $1 / \Mpl(\mu_\s)$, where $\mu_\s$ is the soft scale $\sim \la^2 Q$.
This is because, as we pointed out at the end of \Sec{gravitons}, the ellipses in~\eq{vphi_gauge-trans}---which are $\propto \cO(1 / \Mpl)$ when $\vphi^\s_{\mu\nu}$ is canonically normalized---are NNLP so~\eq{plusplus} is an \emph{exact} infinitesimal diff transformation for a \emph{finite} $\vphi^\s_{\ps_i \ps_i}$ at the LP and also NLP\@.

Finally, let us comment on the apparent nonlocality in $Y_i$, where the soft graviton field is displaced in the $x^{\ps_i}$ direction, a direction \emph{not} allowed for a field to be displaced according to \Sec{nonlocality.1}.
To show that our SCET is a consistent EFT, 
we must demonstrate purely within the EFT, without appealing to the locality of the full theory, that this ``forbidden'' nonlocality is actually a fake.
As the above derivation makes it clear, the origin of $Y_i$ is the field redefinition~\eq{BPS} to achieve a complete manifest separation of soft modes from collinear modes.
We can make the soft graviton couplings manifestly local at the expense of manifest mode separation by simply undoing the field redefinition~\eq{BPS}.
Note that the redefinition~\eq{BPS} can be viewed as just a soft diff gauge transformation on $\Phi_i$ without the accompanying transformation of the soft graviton field,
which has the effect of removing the soft graviton field from a soft diff covariant derivative on $\Phi_i$.
So, undoing~\eq{BPS} puts a derivative of $Y_i$ at where the soft graviton was inside the soft diff covariant derivative.
From the form of $Y_i$, we see that this derivative of $Y_i$ is just $\vphi^\s_{\ps_i \ps_i \!} \del_{\ms_i}$ at the point $x$ of the $\Phi_i(x)$ in question.
This is local. 
Therefore, the ``forbidden'' nonlocality in $Y_i$ is just an artifact of the field redefinition~\eq{BPS}.

Having seen that the nonlocality of $Y_i$ is illusory, 
we should stick to the field-redefined version of $\Phi_i$ (i.e., the one in which $\Phi_i$ is neutral under $G_\s$),
as manifest mode separation is more important than manifest locality from the EFT viewpoint.
(In any case, there are other nonlocal objects in SCET that cannot be field-redefined away.)
There is one thing, however, that still might seem puzzling.
Namely, again in the basis where the field redefinition~\eq{BPS} is undone, 
there are no soft gauge fields appearing in $\cS_\text{hard}$.
But the collinear fields in $\cS_\text{hard}$ are displaced from each other in their respective $x^-$ directions and they are charged under $G_\s$ in \emph{this} basis, so how could $\cS_\text{hard}$ be gauge invariant under $G_\s$ (which we know is nontrivial at the LP)? 
The resolution is that $\cS_\text{hard}$ is actually local from the viewpoint of soft gauge transformations because the physical length scale of nonlocality in $\cS_\text{hard}$ is only $\sim Q^{-1}$ as we discussed in \Sec{nonlocality.1}, 
while the shortest wavelengths of the Fourier modes in soft gauge transformations $U_\s(x) \in G_\s$ are $\sim \la^{-2} Q^{-1}$, much larger than $Q^{-1}$.
Therefore, at the LP and also NLP, soft gauge transformations effectively act as global transformations for $\cS_\text{hard}$.
And since $\cS_\text{hard}$ respects global symmetry, it is also invariant under $G_\s$ at the LP and NLP\@.
Nontrivial effects will arise only if we proceed to NNLP,
which is a conceptually interesting and important problem that is beyond the scope of this paper.
Thus, the ``smearing'' nonlocality discussed in \Sec{nonlocality.2} actually plays a role in making the theory consistent.

With $Y_i$ included in $\cS_\text{hard}$,
we now have all LP interactions of soft gravitons,
which completes the construction of $\cS_\text{hard}$ at the LP\@.
The exponential~\eq{softWilson} is often referred to as \emph{a soft gravitational Wilson line} in the literature (usually without the splitting by $\int'$ (see footnote~\ref{fn:splitting})),
which was identified in full-theory analyses by studying amplitudes in the soft graviton limit~\cite{Brandhuber:2008tf, Naculich:2011ry, Akhoury:2011kq, White:2011yy}.
Our EFT derivation above only uses symmetry and power counting without ever looking at diagrams,   
which not only reveals the true meaning of soft Wilson lines as coming from the field redefinition~\eq{BPS} to achieve manifest mode separation, 
but also makes it self-evident that $Y_i$ gives all LP soft graviton couplings that are correct to all orders in the soft gravitational coupling, $1 / \Mpl(\la^2 Q)$,
as we pointed out above.

\subsection{Summary of gravity SCET at the LP}
\label{s.summary:leading_SCET}
The EFT symmetry and power counting have told us, without any actual calculations, that:
\begin{itemize}
\item{There are no LP collinear graviton couplings anywhere in $\cS_\text{eff}$ except for those that may be already present in $\cS_\text{pure}[\Phi_1, \ldots, \Phi_N]$ at the matching.}
\item{All LP soft graviton couplings are in $\cS_\text{hard}[\Phi_1, \ldots, \Phi_N, \Phi_\s]$, where $\cS_\text{hard}[\Phi_1, \ldots, \Phi_N, \Phi_\s] = \cS_\text{pure}[Y_1 \Phi_1, \ldots, Y_N \Phi_N]$.}
\end{itemize}
Recall our convention that a $\Phi_i$ is a scalar under the $n_i$-collinear diff and transforms covariantly under the $n_i$-collinear local Lorentz group. 
So, in particular, 
an $n_i$-collinear graviton must come in one of the forms \eq{Ri}, \eq{Rimunu}, \eq{Rimunurhosigma} to be a $\Phi_i$.
If derivatives are acting on an $n_i$-collinear field, they should be included inside the $\Phi_i$. 
In particular, this means that the derivatives sit to the right of the $Y_i$. 
That is, the derivatives act on the field first and then the $Y_i$ act.
We have noted that, at the LP,  all those derivatives are just ordinary derivatives $\del$, not covariant derivatives, as far as gravity is concerned.
We have also noted that all vierbeins inside $\Phi_i$ (which are put to make it a diff scalar) should be replaced by Kronecker deltas at the LP\@. 
Therefore, there are no LP collinear graviton couplings ``hiding'' inside $\Phi_i$.  

We now have a complete gravity SCET lagrangian at the LP for our target phase space.
Our derivation using power counting and symmetry shows that this lagrangian is correct at the LP to all orders in the soft and collinear gravitational couplings, $1 / \Mpl(\la^2 Q)$ and $1 / \Mpl(\la Q)$, and at any desired fixed orders in the number of loops and full-theory coupling constants (e.g., $1 / \Mpl(Q)$) at matching.

Needless to say, for \emph{amplitudes} (rather than the lagrangian), 
getting all $1 / \Mpl(\la^2 Q)$ and $1 / \Mpl(\la Q)$ dependences at the LP requires evolving $\cS_\text{hard}$ from the hard scale $\sim Q$ down to the physical scale of interest, by using renormalization group (RG) equations calculated within the SCET\@.
What is the physical scale of interest? 
at the LP,
loops correcting the vertices in $\cS_\text{hard}$ are all coming from soft graviton loops (as far as gravity is concerned).
The scale of virtuality in those loop integrals are hence $\sim \la^2 Q$.
So, the physical scale of interest is $\la^2 Q$.
The RG evolution from $Q$ to $\la^2 Q$ resums a series of powers of large logarithms of the ratio of the hard to soft scales, with each logarithm multiplied by a power of the ``coupling constant'' $Q / \Mpl(\la^2 Q)$.
Such resummation is important if $\la$ is so small that the large logarithm $\log(1 / \la^2)$ compensates for the suppression $Q / \Mpl(\la^2 Q)$.
No logarithms of the ratio of the hard to collinear scales exist at the LP because there are no collinear graviton corrections to $\cS_\text{hard}$ at the LP
so the $1 / \Mpl(\la Q)$ dependence we get from matching is actually already correct without RG evolution.

In gravity SCET, no actual calculations are necessary to see that there are no LP couplings of collinear gravitons, thanks to mode separation and the factorized gauge symmetry.
Demonstrating this result in the full theory is extremely nontrivial.
In a full-theory diagram where an $n_i$-collinear graviton is attached to an $n_j$-collinear line with $j \neq i$, it appears that the coupling is ``Bigger-than-Leading'' Power (BLP) because $h^i_{\mu\nu} p_j^\mu p_j^\nu \sim \ds{(n_{\ps_i} \dt n_{\ps_j})^2} / \la \sim 1 / \la$ from the cross-collinear scaling~\eq{cross-collinear} and the collinear graviton scaling~\eq{collinear-h-scaling}.
However, after adding up all diagrams with all possible places that this graviton can be attached to, one finds that all BLP and LP contributions completely cancel out.
For example, the amplitudes in Appendices~\ref{a.Phi4}--\ref{a.ssff} exhibit these dramatic cancellations if calculated in the full theory.
Such cancellations can be demonstrated in the full theory in a full generality through a careful combinatorial analysis combined with Ward identities~\cite{Akhoury:2011kq}, while our gravity SCET lagrangian simply does not have any BLP or LP couplings of collinear gravitons from the outset.
Thus, our EFT passes the test of manifest power counting in the sense described in \Sec{manifest}.

\subsection{Soft/collinear theorems for gravity at the LP}
\label{s.soft-collinear-theorems}
The LP soft theorem for gravity~\cite{Weinberg:1965nx, Gross:1968in}\cite{Naculich:2011ry, Akhoury:2011kq, White:2011yy} is completely self-evident in the EFT at the lagrangian level, without any need for analyzing amplitudes or diagrams.
The structure of $\cS_\text{hard}$ derived in \Sec{leading_order}---a string of $Y_i$'s acting on $\cL_\text{pure}[\Phi_1, \ldots, \Phi_N]$---already has the form of ``a universal soft factor $\otimes$ the hard interaction''.
(And recall that there are no LP soft couplings in $\cS_\text{full}[\Phi_\s]$.)
We arrived at this structure of $\cS_\text{hard}$ only based on the factorized gauge symmetry and power counting.
This situation is analogous to the demonstration of the soft theorem for spin-1 gauge theories by SCET~\cite{Larkoski:2014bxa}. 
This not only shows a power of EFTs but also provides us with an understanding of soft theorems in terms of purely long-distance properties of the theory,
as they should be understood.

The collinear theorem at the LP~\cite{Weinberg:1965nx}\cite{Akhoury:2011kq} is even more trivial. 
There are simply no collinear graviton couplings at the LP\@.
Physically, there is no such thing as a ``graviton jet''.
This does not mean that there should not be any gravitons in the scattering process at the LP\@. 
A collinear sector may consist of a graviton in $\cS_\text{pure}$ in the purely hard limit, 
but it is not possible for any \emph{other} graviton that is collinear to that graviton to exist at the LP\@.
Again, we only need symmetry and power counting to see all this,
which is very nontrivial from the diagrammatic perspective of the full theory as
we already noted at the end of \Sec{summary:leading_SCET}. 

Finally, the fact that collinear gravitons manifestly decouple \emph{in the lagrangian} in the $\ds{\la \to 0}$ limit immediately implies the absence of collinear IR divergences from gravitational interactions in all processes in our target phase space. 
There is no need to analyze diagrams, because the decoupling already occurs at the lagrangian level so we cannot even \emph{conceive} diagrams that might be potentially  collinear divergent in gravity SCET\@.
Put it another way, in any individual diagram of gravity SCET, every time we detach a collinear graviton line from a vertex, the $\la$ dimension of the diagram decreases by at least one. Repeating such detaching procedure, we eventually arrive at a diagram with no collinear gravitons, which has no collinear IR divergence due to gravity. The original diagram we started with has a $\la$ dimension higher than this, so it is obviously collinear IR finite. 
This should be contrasted to the QCD SCET situation in which the collinear gauge interactions are LP and thus fully remain in the effective lagrangian in the $\ds{\la \to 0}$ limit, leading to collinear divergences.

On the other hand, soft graviton couplings from $Y_i$ are LP so we should expect soft IR divergences in gravity SCET from those soft graviton couplings.
However, unlike in the full theory, the source of soft divergences is completely isolated in the EFT\@. Namely, soft divergences only appear in the matrix element of the product of $Y_i$'s from $\cS_\text{hard}$.
(Recall that soft graviton couplings in $\cS_\text{full}[\Phi_\s]$ are suppressed at least by $\la^2$ and thus completely decouple in the $\ds{\la \to 0}$ limit.)

\section{Gravity SCET at the Next-to-Leading Power (NLP)}
\label{s.order-lambda}
%

\subsection{Self-$n_i$-collinear gravitational interactions at the NLP}
The argument in \Sec{no_collinear_gravity_at_leading_order} suggests that there should be $\cO(\la)$ couplings of $n_i$-collinear gravitons to $n_i$-collinear particles in $\cS_\text{full}[\Phi_i]$.
We can obtain all such couplings by expanding $\cS_\text{full}[\Phi_i]$ to $\cO(\la)$
by following the power counting rules described in \Secs{modeseparation}, \ref{s.gravitons}, and \ref{s.matter}. 

Note that those couplings are suppressed not only by $\la$ but also by $Q / \Mpl$.
We only care about $\la$ expansion in this paper, 
but we would like to point out that the only invariant energy scale in $\cS_\text{full}[\Phi_i]$ is $\la Q$, not $Q$. 
Therefore, the validity of $\cS_\text{full}[\Phi_i]$ as an EFT in powers of $1 / \Mpl$ only requires $\la Q \ll \Mpl$.
The hard scale $Q$ itself can be much larger than $\Mpl$ 
as far as $\cS_\text{full}[\Phi_i]$ is concerned.
This would be an important point if we would like to construct a gravity SCET for scattering processes in the Regge limit.

\subsection{Collinear gravitational Wilson lines at the NLP}
\label{s.collinear-wilson-lines}
As we already pointed out in \Sec{collinear_invariance}, 
an $n_i$-collinear operator $\Phi_i$---which may correspond to a matter particle or to a graviton in the form of the covariant gravitational objects~\eq{Ri}, \eq{Rimunu}, \eq{Rimunurhosigma}---does transform at the NLP under the $n_i$-collinear diff$\times$Lorentz gauge transformations~\eq{collinear-matter-collinear-transform}. 
Therefore, we must make them gauge invariant also at the NLP so that we can put them in $\cS_\text{hard}$.
As it was realized in the original QCD SCET, 
the marvelous trick is to exploit the nonlocality of SCET discussed in~\Sec{nonlocality.1} and use nonlocal objects, namely Wilson lines, to accomplish the desired invariance.

\subsubsection{Collinear gravitational Wilson lines for collinear local Lorentz groups}
\label{s.WL:col:Lorentz}
Like the usual spin-1 gauge theory case, 
a Wilson line along a path $\cP$ for the local Lorentz gauge group acting on representation $r$ is given by 
\beq
W_r[\cP] \equiv 
\hat{\cP} \exp\lt[ -\fr12 \!\int_\cP \dd z^\mu \, \ga_{\mu \al\be}(z) \, \Sg_r^{\al\be} \rt],
\eeq
where the line integral is taken along the path $\cP$ with $\hat{\cP}$ indicating path-ordering, while $\Sg_r^{\al\be}$ ($\al, \be = 0, \ldots, 3$) are the Lorentz generators for the representation $r$ satisfying the Lorentz algebra~\eq{lorentzalgebra}.
Finally, $\ga_{\mu \al\be}$ is the spin connection, where our convention is such that the local-Lorentz covariant derivative is given by $\DD_{\mu} = \del_\mu + \frac12 \ga_{\mu\al\be} \Sg^{\al\be}$. 
We then have
\beq
\ga_{\mu\al\be} 
= -\fr12 \Bigl( 
\del_\mu \vphi_{\al\be} + \del_\al \vphi_{\mu\be} + \del_\al \vphi_{\be\mu} 
- ( \al \lr \be) \Bigr)
+ \cO(\vphi^2)
\,.\eql{spin_connection}
\eeq
Now, we want one end of the Wilson line to extend to an infinity so that we can render the field it acts on to be gauge invariant.
Since fields are only allowed to be displaced in the $x^{\ms_i}$ coordinate in SCET, 
the Wilson line should be given by
\beq
W^i_r(x) \equiv 
\hat{\cP} \exp\lt[ -\fr12 \int'\!\! \dd s \, \ga^{(i)}_{\ms_i \, \al\be}(x + s n_{\ms_i}) \, \Sg_r^{\al\be} \rt],
\eql{col_Wilson_Lorentz_future}
\eeq
where $\int'$ is defined in~\eq{Is} (also see footnote~\ref{fn:splitting}).
The argument that led to the use of $\int'$ works just in the same way except that it is $p^{\ms_i}$, not $p^{\ps_i}$, that picks up a $+\I\ep$ here.
The extra superscript~${}^{(i)}$ of $\ga^{(i)}_{\mu\al\be}$ indicates that this spin connection is made of the $n_i$-collinear graviton field $\vphi^i_{\mu\nu}$, not the full $\vphi_{\mu\nu}$, 
because this Wilson line is for the $n_i$-collinear local Lorentz gauge group.
Then, by construction, the product $\ds{W^i_r(x) \, \Phi_i^{(r)\!}(x)}$ is invariant under the $n_i$-collinear local Lorentz gauge group.
Note that the product still transforms covariantly as the representation $r$ under the \emph{global} Lorentz group.
Because of the factorized gauge symmetry structure~\eq{factorized_symm},
$W_r^i$ can only act on an $n_i$-collinear operator, never on an $n_j$-collinear operator with $j \neq i$ or on a soft operator.

Let us now power-count the exponent of the Wilson line~\eq{col_Wilson_Lorentz_future}.
We expect it to be NLP, but NLP does not mean that the exponent is $\cO(\la)$ because $W^i_r$ have different components and they scale differently.
This remark also applies to the different components of $\Phi_i^{(r)}$ that $W^i_r$ acts on.
Again, to avoid this complication, let's boost in the $\vec{n}_{\ps_i}$ direction by a rapidity of $\sim \log\la$ such that $n_i$-collinear momenta now scale as $\sim (\la, \la, \la)_i$.
This is a global Lorentz boost so the spin connection $\ga_{\mu\al\be}$ also transforms covariantly as everybody else.
In this boosted frame, all components of $\Phi_i^{(r)}$ scale with a common power of $\la$, and all components of $\vphi^i_{\mu\nu}$ scale as $\sim \la$. 
Now, let's look at the exponent of~\eq{col_Wilson_Lorentz_future}.
Since $s$ parametrizes the $x^{\ms_i}$ coordinate, we have $\dd s \sim \dd x^{\ms_i} \sim \la^{-1}$.
The bounds of the integration, $0$ and $\infty$, do not scale with $\la$.
For the spin connection, according to~\eq{collinear-h-scaling},
the terms explicitly shown in~\eq{spin_connection} scale as $p_\mu p_{\al} p_{\be} / \la$ with an $n_i$-collinear momentum $p$, and thus as $\sim \la^2$.
The terms implicit in~\eq{spin_connection} are quadratic or higher in $\vphi^i_{\mu\nu}$ and hence suppressed by an extra power or powers of $\la$, so they are subleading to the explicitly shown terms.
Combining all the pieces together, we see that, in the boosted frame, the exponent of~\eq{col_Wilson_Lorentz_future} scales as $\sim \la^{-1} \la^2 = \la$.
We therefore must Taylor-expand $W^i_r$ to $\cO(\la)$ and truncate the higher order terms 
in order to make power counting manifest in the effective action and not to include higher order terms we have no right to keep. 
Once Taylor-expanded and multiplied by $\Phi_i^{(r)}$ on the right, 
we can boost back to the original frame, 
which does not change the fact that the first-order term is suppressed by $\la$ compared to the zeroth-order term, thanks to Lorentz covariance of each term. 

To summarize, for constructing a gravity SCET to the NLP, the $n_i$-collinear Lorentz Wilson line is given by 
\beq
W^i_r(x) =
1 -\fr12 \int'\!\! \dd s \, \ga^{(i)}_{\ms_i \, \al\be}(x + s n_{\ms_i}) \, \Sg_r^{\al\be}
\,.\eql{LorentzWilson}
\eeq
%

\subsubsection{Collinear gravitational Wilson lines for collinear diff groups}
\label{s.WL:col:diff}
By multiplying a collinear Lorentz Wilson line~\eq{LorentzWilson}, 
we have made every $\Phi_i^{(r)}$ in $\cS_\text{hard}$ invariant under all collinear Lorentz gauge groups.
Let $\wtd{\Phi}_i$ be such a collinear-Lorentz invariant object.
As we already pointed out below~\eq{collinear-matter-collinear-transform}, 
such $\wtd{\Phi}_i$ is no longer invariant at the NLP under $n_i$-collinear diff transformations  but transforms as
\beq
\wtd{\Phi}_i \too \wtd{\Phi}_i + \xi_i^{\mu} \del_\mu \wtd{\Phi}_i
\,,\eql{col_diff_trans}
\eeq
where the last term above scales as $\sim \la \wtd{\Phi}_i$.
So, in order to make $\cS_\text{hard}$ fully gauge invariant also at the NLP, 
we must find an object $V_i$ such that $V_i \wtd{\Phi}_i$ is invariant under the $n_i$-collinear diff group to the NLP\@.
Here, we do not need a superscript ${}^{(r)}$ for $\wtd{\Phi}_i$ nor $V_i$ because the transformation~\eq{col_diff_trans} is independent of $r$.
So, we already know that collinear graviton couplings from $V_i$ are spin independent.

Again thanks to the nonlocality of SCET, we can find $V_i$.
Since \eq{col_diff_trans} is an infinitesimal translation in spacetime, 
it is natural to guess that $V_i$ must schematically have the form $\sim 1 + \Ga$ when expanded to the NLP, 
where $\Ga$ is the Christoffel connection. 
To determine the precise form, note that $\Ga^\mu_{\nu\rho}$ transforms under a general infinitesimal diff transformation $\de x^\mu = -\xi^\mu(x)$ as
\beq
\de\Ga^\mu_{\nu\rho} 
= 
\del_\nu\del_\rho\xi^\mu
-(\del_\al \xi^\mu) \Ga^\al_{\nu\rho}
+(\del_\nu \xi^\al) \Ga^\mu_{\al\rho}
+(\del_\rho \xi^\al) \Ga^\mu_{\nu\al}
+\xi^\al \del_\al \Ga^\mu_{\nu\rho}
\,,\eql{Chris_diff}
\eeq
where higher powers of $\xi$ are neglected but the exact dependence on the graviton field at $\cO(\xi)$ is kept.
Now, for the $n_i$-collinear sector, the collinear scalings~\eq{collinear-h-scaling} and~\eq{collinear-xi-scaling} tell us that the first term of the right-hand side of~\eq{Chris_diff} scales with an $n_i$-collinear momentum $p$ as
\beq
\del_\nu\del_\rho\xi^\mu \sim p_\nu p_\rho p^\mu / \la
\,,
\eeq
while all other terms on the right-hand side scale as
\beq
p_\al p^\mu p^\al p_\nu p_\rho / \la^2 
\sim p^\mu p_\nu p_\rho
\,.
\eeq
So, at the leading nontrivial order in $\la$, the diff transformation~\eq{Chris_diff} simply reduces to
\beq
\de\Ga^\mu_{\nu\rho} 
= 
\del_\nu\del_\rho\xi^\mu
\,.\eql{Chris_diff_leading}
\eeq
Thus, the integral $-\int \dd x^\nu \, \dd x^\rho \, \de\Ga^\mu_{\nu\rho} \,\del_\mu$ will give us $-\xi^\mu \del_\mu$ and we can use this to cancel the unwanted term in~\eq{col_diff_trans}. 
The integration path should be taken in the $x^{\ms_i}$ direction, the allowed direction for fields to be displaced in the $n_i$-collinear sector.
And we must take care of the convergence of the integrals consistently with the $+\I\ep$ prescription as we did for $Y_i$ and $W_r^i$. 
So, we find that $V_i$ should be given by
\beq
V_i(x) 
= 1 
- \int_{-\infty}^0 \!\!\! \dd s' \!\int_{-\infty}^{s'} \!\!\! \dd s \>
\Ga^{\mu(+)}_{\ms_i \ms_i \!}(x + s n_{\ms_i}) \, \del_\mu
- \int^{\infty}_0 \!\! \dd s' \!\int^{\infty}_{s'} \!\! \dd s \>
\Ga^{\mu(-)}_{\ms_i \ms_i \!}(x + s n_{\ms_i}) \, \del_\mu
\,,\eql{diffWilson}
\eeq
where $\Ga^{\mu(+)}_{\ms_i \ms_i \!}$ and $\Ga^{\mu(-)}_{\ms_i \ms_i \!}$ denote the positive and negative frequency parts of $\Ga^{\mu}_{\ms_i \ms_i \!}$, respectively. (Unfortunately, the $\int'$ notation we used for $Y_i$ and $W_r^i$ does not work here, but the comment in footnote~\ref{fn:splitting} also applies to $V_i$.)

Let's quickly power-count $V_i(x)$.
The derivative $\del_\mu$ in \eq{diffWilson} acts on an $n_i$-collinear operator so $\del_\mu \sim p_\mu$ with an $n_i$-collinear momentum $p$.
The $s$ and $s'$ variables are shifts in the $x^{\ms_i}$ coordinate, which is $\sim \la^0$, so they do not scale. Finally, $\Ga^\mu_{\ms_i \ms_i}$ scales as $\sim p^\mu p_{\ms_i} p_{\ms_i} / \la \sim p^\mu / \la$ as we discussed above.
We thus see that $V_i = 1 + \cO(\la)$, as it should be.

\subsection{Comments on nonlocal ``dressing'' of operators}
All of our Wilson lines, $Y_i(x)$, $W_r^i(x)$, and $V_i(x)$, can be thought of as particular realizations of the notion of ``dressing'' discussed in \cite{Donnelly:2015hta, Donnelly:2016rvo}.
The motivation of~\cite{Donnelly:2015hta, Donnelly:2016rvo} for dressing is to find diff invariant observables in quantum gravity that would return to the usual local operators in the limit of no gravity, 
and they discuss various possible forms of diff invariant dressing of local operators, 
including the ones very similar to, but still different from, our $V_i(x)$. 
In our SCET, we derived the specific forms of dressing via $Y_i$, $W_r^i$, $V_i$ from the factorized gauge symmetry~\eq{factorized_symm},
which is ``merely'' an \emph{effective} gauge symmetry at \emph{long} distances and thus makes no reference to the ultimate nature of quantum gravity beneath the Planck length.

\subsection{Soft graviton couplings at the NLP}
\label{s.soft_lambda}
Here, we would like to show, again just using symmetry and power counting, that there are no NLP couplings of soft gravitons to anything.

First, as already pointed out in \Sec{soft_selfcouplings_order_one}, 
there are no $\cO(\la)$ soft graviton couplings to soft particles, including soft gravitons themselves.
Next, for NLP couplings of soft gravitons to $n_i$-collinear particles,
we need to expand the infinitesimal soft diff$\times$Lorentz transformation~\eq{collinear-matter-soft-transform} to the NLP: 
\beq
\Phi_i \too 
\Phi_i 
+ \xi^\s_{\ps_i} \del_{\ms_i} \Phi_i
+ \xi^\s_{\pp_i} \!\dt\, \del_{\pp_i} \Phi_i
\,,\eql{Phi_i-subleading-soft-transform}
\eeq
where the second-to-last term is what we already had at the LP
while the last term is a new term. 
The new term is suppressed by $\la$ compared to the other terms, 
because $\ds{\xi^\s_{\pp_i} \sim \la^0}$ and $\ds{\del_{\pp_i} \Phi_i} \sim \la \Phi_i$.
The transformation~\eq{Phi_i-subleading-soft-transform} is still spin independent,
which is why we have suppressed the superscript ${}^{(r)}$ of $\Phi_i^{(r)}$.

Now, it might appear that there should be NLP soft graviton couplings to collinear particles because $\Phi_i$ does seem to transform at the NLP\@.
However, the last term in~\eq{Phi_i-subleading-soft-transform} can actually be eliminated by exploiting RPI we mentioned in \Sec{cross-collinear-RPI}, namely, the invariance of the theory under the redefinition (or reparametrization) of the basis vectors $n_{\pm_i}$. Now, because we have already established that there are no collinear gravitational interactions at the LP, there is only one $n_i$-collinear particle%
\footnote{Note that this ``particle'' may be a stream of nearly exactly collinear, non-gravitationally splitting particles as discussed along with the last assumption stated in \Sec{target}.}
in the $n_i$-collinear sector at the LP\@.
So, we can always redefine the basis vectors $n_{\pms_i}$ by adding to them some vectors lying in the $\perp_i$ plane such that the new $\perp_i$ component of the momentum of this particle vanishes exactly.
In such basis of $n_{\pms_i}$, 
the last term in~\eq{Phi_i-subleading-soft-transform} is zero
and hence the LP expression~\eq{Phi_i-leading-soft-transform} we used in \Sec{leading_soft_couplings_to_collinear_matter} now becomes also correct at the NLP\@.
The transformation of the soft graviton field~\eq{plusplus} we used there also remains unchanged at the NLP, 
because that is just the $+_i +_i$ component of the transformation law~\eq{vphi_gauge-trans}, which is also correct at the NLP as we discussed at the very end of \Sec{gravitons}.
Therefore,
once $n_{\pms_i}$ is chosen for every $i$ such that the $\perp_i$ component of the momentum of $\Phi_i$ vanishes,  
the soft graviton couplings from $Y_i$ we obtained in \Sec{leading_soft_couplings_to_collinear_matter} become also correct at the NLP\@.

\subsection{Other sources of NLP contributions}
\label{s.other_lambdas}
There are other possible sources of NLP contributions to $\cS_\text{hard}$:%
\begin{itemize}
\item[(a)]{An insertion of $R^i_{\ms_i \ms_i}$ or $R^i_{\ms_i \pp_i \ms_i \pp_i}$ into an otherwise LP operator.
Needless to say, we now have an independent integration over the $x^{\ms_i}$ coordinate of the inserted field, in addition to the $x^{\ms_i}$ integration for the existing $n_i$-collinear field. The NLP operator thus constructed has its own Wilson coefficient with dependence on the $x^{\ms_i}$ coordinate of the inserted field.
Of course, each of these $R^i_{\ms_i \ms_i}$ and $R^i_{\ms_i \pp_i \ms_i \pp_i}$ must be multiplied by $Y_i$ to take into account $G_\text{s}$.
}
\item[(b)]{The replacement of the leading component of an $n_i$-collinear field by a subleading component of the same field.
An example is the replacement of $P_i^+ \psi^i$ with $P_i^- \psi^i$ for a spinor $\psi^i$.
}
\item[(c)]{The 2-graviton terms in~\eq{Ri}, \eq{Rimunu}, or \eq{Rimunurhosigma}, if the LP operator already has an $n_i$-collinear graviton in one or more of those three forms.
}
\item[(d)]{At the NLP, we must be careful about our conventions that $\Phi_i^{(r)}$ should include all derivatives acting on the ``elementary'' field in question 
and that $\Phi_i^{(r)}$ should transform covariantly under the $n_i$-collinear Lorentz group.
Since the field transforms at the NLP under the $n_i$-collinear Lorentz group, 
those derivatives must be $n_i$-collinear covariant derivatives in terms of the $n_i$-collinear spin connection, $\ga^{(i)}_{\mu\al\be}$, 
which contain NLP $n_i$-collinear graviton couplings.
Here, it suffices to only include the one-graviton terms explicitly shown in~\eq{spin_connection} because terms quadratic or higher in the graviton field are NNLP or higher.
}
\item[(e)]{Similarly, if $\Phi_i^{(r)}$ contains vierbeins in it, 
they may contain NLP couplings. 
See the relevant discussion in \Sec{collinear_invariance}.
Again, we only need to expand the vierbeins to first order in the graviton field for NLP couplings.}
\end{itemize}
Among these, the NLP couplings from (c)--(e) are completely fixed by the diff$\times$Lorentz gauge invariances and hence do not introduce any new parameters.
The couplings via (b) are fixed by RPI and hence again introduce no new parameters. 
To see this, note that $n_i$-collinear RPI may be viewed as $n_i$-collinear Lorentz invariance with the lightcone basis vectors $n_{\pm}$ being treated as spurions~\cite{Manohar:2002fd}. 
In other words, an expression is reparametrization invariant if it does not refer to $n_\pm$.
So, for example, when we replace a $P_i^+ \psi^i$ with a $P_i^- \psi^i$, RPI tells us that the latter should come with exactly the same coefficient as the former,
because the only Lorentz covariant combination of $P_i^+ \psi^i$ and $P_i^- \psi^i$ without referring to $n_\pm$ is $P_i^+ \psi^i + P_i^- \psi^i = \psi^i$.
Similar arguments apply to fields with higher spins.

It is also quite possible that the operators from~(a) are also fixed by RPI, 
or local Lorentz symmetries with the spurions $n_\pm$. 
Indeed, the concrete examples discussed in Appendices~\ref{a.Phi4}--\ref{a.ssff} suggest that this might be the case.
However, (dis-)proving the case requires an understanding of how RPI should be modified by the curvature of spacetime, at least to the NLP if not to all orders.
Identifying the group of such RPI transformations and exploring its full implications clearly constitute an interesting and important problem that deserves its own separate study~\cite{progress}.
Another likely possibility is that the operators from~(a) are not fixed but there are universality classes of gravity SCETs and the theories~\ref{a.Phi4}--\ref{a.ssff} belong to the same universality class. This would be also extremely interesting. 

Next, let's examine the fact that, in terms of power counting alone, 
the replacement of a $\del_{\ms_i}$ with a $\del_{\perp_i}$ in $\cS_\text{hard}$ would yield an NLP contribution.
However, RPI tells us that these contributions vanish in the basis described in \Sec{soft_lambda} in which the $\perp_i$ component of the momentum of $\Phi^{(r)}_i$ vanishes.
One might also wonder that, if $\cS_\text{hard}$ already had a $\del_{\pp_i}$ at the LP, the subleading $\cO(\la^2)$ fluctuations in the $\del_{\perp_i}$ would also yield NLP contributions (see footnote~\ref{fn:spiritviolation}).
However, this is actually not possible. If we could eliminate terms with $\del_{\pp_i}$ by reparametrizing $n_{\pm_i}$, that would mean that there should be other terms in $\cS_\text{hard}$ that would produce $\del_{\pp_i}$ upon such reparametrization.
But reparametrization in the $\perp_i$ plane is $\cO(\la)$.
Hence, terms with $\del_{\pp_i}$ cannot be LP but must be NLP at most, 
so the subleading $\cO(\la^2)$ fluctuations in the $\del_{\perp_i}$ would actually be NNLP effects.        

One might wonder if there are more subtle NLP contributions.
For example, rather than displacing $x^\mu$ straight to $\ds{x^\mu + s n_{\ms_i}^\mu}$ for the nonlocal integrations in $\cS_\text{hard}$, shouldn't we displace it along some curved path? 
We don't have to worry about this.
Since the path is straight at the LP, if it is curved at the NLP, 
it is in the form of a straight LP path plus an NLP deviation.
So, the NLP deviation can be just Taylor-expanded around the straight path.
For non-scalar fields, NLP deviations in the paths will be also accompanied by NLP twists in the directions of the fields, which can again be taken into account by derivatives and Lorentz generators acting on the fields.  
Therefore, the curved path effects, if any, are already implicitly included by writing down all possible derivatives and tensor structures for operators in $\cS_\text{hard}$. 
It is tempting to wonder if the contributions of type~(a) could arise from describing curved paths in an RPI manner.

Possible NLP pieces in the cross-collinear scaling~\eq{cross-collinear} can also be likewise taken into account by writing down all possible derivatives and tensor structures.
Thus, the right-hand side of~\eq{cross-collinear} can be regarded as strictly $\la^0$ without subleading $\cO(\la)$ pieces.
Similarly, we do not have to worry about subleading corrections to cross-collinear contractions of bosonic fields as well as those of fermions, \eq{fermion-cross-collinear} and \eq{gravitino-cross-collinear}.

\subsection{Recipe for constructing a gravity SCET lagrangian to the NLP}
\label{s.recipe}
Here's a complete recipe for how to construct a gravity SCET that is correct to the NLP for scattering processes in our target phase space:%
\begin{itemize}
\item[(O)]{Expand all gravitational interactions in $\cS_\text{full}[\Phi_i]$ to $\cO(\la)$, following the power counting rules in \Secs{modeseparation}, \ref{s.gravitons}, and \ref{s.matter}. Gravitational interactions in $\cS_\text{full}[\Phi_\s]$ can be completely discarded as they are all $\cO(\la^2)$.} 
\item[(I)]{From purely hard matching, determine $\cS_\text{pure}[\Phi_1, \ldots, \Phi_N]$ at the \emph{LP}\@.}
\item[(II)]{Replace each $\Phi_i^{(r)\!}(x)$ in $\cS_\text{pure}[\Phi_1, \ldots, \Phi_N]$ by the product $\ds{Y_i(x) \, V_i(x) \, W^i_r(x) \, \Phi_i^{(r)\!}(x)}$. 
Recall our convention that $\Phi_i^{(r)\!}(x)$ must be made a diff scalar by vierbeins 
and it should also include all derivatives acting on the field in question.}
\item[(III)]{Get NLP couplings of types~(b)--(e) of \Sec{other_lambdas}.}
\item[(IV)]{Add NLP operators of type~(a) of \Sec{other_lambdas} and determine their Wilson coefficients by matching at the NLP\@.}
\item[(V)]{Calculate renormalization group equations using the SCET thus obtained. 
(Of course, this must be done at the loop order higher by one than the matching loop order, as usual.) 
Then, run $\cS_\text{hard}(\mu)$ from the hard scale $\mu \sim Q$ down to the soft scale $\mu \sim \la^2 Q$ to resum large logarithms of the ratio of the hard to soft scales.}
\end{itemize}
Steps (I) and (IV) can be done at any desired fixed order in the number of loops and full-theory coupling constants, in particular, $1 / \Mpl(Q)$.
Then, steps (O) and (II)--(V) will provide amplitudes with all LP and NLP contributions from collinear and soft gravitons, which are correct to all orders in $1 / \Mpl(\la Q)$ and $1 / \Mpl(\la^2 Q)$.
Regarding step (V), note that there are no collinear logarithms to resum---as far as gravity is concerned---as there are no LP collinear graviton interactions.
Needless to say, zero-bin subtraction must be performed to remove the double-counting as we mentioned in~\Sec{fact_symm}.

\subsection{Soft/collinear theorems for gravity at the NLP}
From our discussions above, we see that the LP soft theorem is also correct at the NLP, provided that we regard the collinear graviton couplings of type~(a) of \Sec{other_lambdas} as belonging to the ``hard'' part of the soft theorem. 
But it should be emphasized that this conclusion relies on the use of RPI discussed in \Sec{soft_lambda}, which in turn relies on the assumption of our target phase space that the angles of collinear splittings due to non-gravitational interactions are hierarchically different from those due to gravity.
Relaxing this assumption and studying how the soft theorem should be modified at the NLP is an important and interesting problem.

In contrast, the decoupling of collinear gravitons no longer holds at the NLP\@.
Namely, we have a new collinear graviton in one of the collinear sectors.
However, in order to make a collinear \emph{theorem,} that is, to make some universal statements, we need to be able to say that the contributions from operators of type~(a) of \Sec{other_lambdas} are fixed by RPI or something, or show that there are more than one universality class and classify all possible universality classes. 
We defer this very important and interesting question to our future publication~\cite{progress}.

However, even without any universal theorem, the structures of LP and NLP operators dictated by the symmetry and power counting of the EFT dramatically simplify the calculations of NLP contributions to amplitudes. 
In the EFT, we only need to calculate one term from each of NLP operators of type~(a) of \Sec{other_lambdas} to determine its Wilson coefficient, and then we will have the entire NLP amplitudes. 
In contrast, in the full theory, one needs to calculate every single NLP term, which is possible only after carefully and laboriously expanding each vertex and each propagator of each diagram down to NLP starting, typically, at BLP\@.
One can see this contrast even in the simple examples in Appendices~\ref{a.Phi4}--\ref{a.ssff}.

\subsection{Comparison with the literature}
\label{s.comparison}
Ref.~\cite{Beneke:2012xa} made the first attempt to construct SCET for gravity at the LP, where the main thrust of the paper is to provide a SCET-like demonstration of the decoupling of BLP collinear graviton couplings, although it also discusses the soft sector.
However, while their decoupling argument itself may be valid,   
the structure of their effective lagrangian prior to the decoupling, \emph{viewed as a SCET,} has the following problems. 
They introduce a collinear Wilson line of the form (in our convention):
\beq
\widetilde{V}(x^+) = \exp\lt[ \fr12 \int_{-\infty}^0 \!\! \dd s \, h_{--}(x^+ \!+ s) \, \del_+ \rt].
\eeq
As can be seen from the absence of a Lorentz generator above, this Wilson line is for the diff part of diff$\times$Lorentz, like our $V_i$.
In Ref.~\cite{Beneke:2012xa}, $\widetilde{V}$ is multiplied onto fields in \emph{different} collinear sectors from the sector that the $h_{--}$ inside $\widetilde{V}(x)$ belongs to, thereby describing BLP couplings of collinear gravitons in our language. 
In fact, the core of the paper is to show that $\widetilde{V}$, and hence the BLP couplings, can be completely removed from the lagrangian by a coordinate transformation or equivalent field redefinition.
In our viewpoint, there are four problems here.
First, unlike our $V_i$, $\widetilde{V}$ does not transform under diff as a Wilson line should. One can check this explicitly, but also recall our analysis of diff invariance in \Sec{leading_soft_couplings_to_collinear_matter} that determines the diff Wilson line uniquely to be $V_i$, not $\widetilde{V}$. 
Therefore, neither the form of $\widetilde{V}$ itself nor the rule for how $\widetilde{V}$ should enter the lagrangian is based on symmetry but one needs to look at full-theory diagrams to see those.
Second, even if it did have the right transformation property as a Wilson line within the collinear sector that $\widetilde{V}$ belongs to, the factorized gauge symmetry would forbid $\widetilde{V}$ to be multiplied onto a field in a different collinear sector.
So, there would be no need to go through a coordinate transformation or field redefinition to decouple $\widetilde{V}(x)$ from the lagrangian, because it could not be written down in the first place.
Third, given that it had been already known from full theory analyses~\cite{Akhoury:2011kq} that BLP and LP collinear graviton couplings should always cancel, 
the right EFT with manifest power counting and the right effective symmetries should 
exhibit their absence from the outset.
Indeed, in our gravity SCET, the largest couplings of collinear gravitons ever permitted by symmetry are NLP\@. 
In contrast, the theory of Ref.~\cite{Beneke:2012xa} apparently permits BLP couplings via $\widetilde{V}$ to be written down, although they have clear arguments for why there must exists a coordinate transformation that removes such couplings,
and also show an explicit expression for field redefinition that removes $\widetilde{V}$.
While those arguments and field redefinition themselves may be valid, 
they do not constitute a \emph{SCET} demonstration of the decoupling.
Finally, while our analysis shows the absence of both BLP and LP couplings, Ref.~\cite{Beneke:2012xa} does not discuss the decoupling of LP collinear graviton couplings, which in particular requires to discuss not only $h_{--}$ but also $h_{-\perp}$ components.

On the other hand, for the soft Wilson line $Y_i$, we agree on its final form with Ref.~\cite{Beneke:2012xa} as well as the older, full-theory studies~\cite{Naculich:2011ry, Akhoury:2011kq, White:2011yy}. In Ref.~\cite{Beneke:2012xa}, after the soft Wilson line is introduced it is verified that it has the right transformation property as a Wilson line. Our more deductive derivation in terms of symmetry also makes clear the uniqueness of its form as well as the fact that it does not get modified at the NLP as discussed in~\Sec{soft_lambda}.

\section{Conclusions and future directions}
In this paper, we have identified fundamental building blocks of gravity SCET and laid out a procedure for writing down the effective lagrangian for any given full theory
at the leading power and the next-to-leading power 
for processes that belong to our target phase space.
In particular, we identified basic building blocks of the EFT, the most notable of which being the soft, collinear Lorentz, and collinear diff Wilson lines: $Y_i$, $W_r^i$, and $V_i$, respectively.
The soft theorem and the decoupling of collinear gravitons at the leading power are structurally manifest at the lagrangian level in the gravity SCET\@. 
Permeating and underlying all of our analyses are mode separation, symmetry---especially the factorized gauge symmetry---and power counting.
(But one should recall that factorized gauge symmetry is a consequence of mode separation, which we need for manifest power counting, 
which is compulsory for an EFT to systematically control the kinematics of its target phase space.) 
All results and claims in this paper are derived from those principles without recourse to diagrammatic analyses. 
The gravity SCET lagrangian thus constructed is now ready for perturbative calculations with effective symmetries and power counting maximally manifest unlike calculations in the full theory, which greatly facilitate the calculations.

There are some obvious variations of our gravity SCET\@.
First, the target phase space may be modified.
As we alluded to in Introduction, if we relax our assumptions that all particles have no or negligible mass compared to the soft scale $\la^2 Q$, 
we will need to adapt our path of building gravity SCET for massive particles.
Or, if we relax the assumption that non-gravitational collinear splittings occur with a much larger or smaller ``$\la$'' than the $\la$, 
we will need to simultaneously include all gauge groups in the factorized effective symmetry and introduce corresponding collinear and soft Wilson lines for all of them.

Being an EFT for highly energetic collinear particles, 
SCET can also be useful for studying the amplitudes of extremely energetic forward scattering (the Regge limit).
The Regge limit may give us another interesting channel toward understanding gravity~{\cite{tHooft:1987vrq, Verlinde:1991iu, Giudice:2001ce, Giddings:2007bw, Giddings:2010pp, Melville:2013qca, Akhoury:2013yua, Luna:2016idw}. 
In QCD SCET, it has been established~\cite{Fleming:2014rea, Donoghue:2014mpa, Rothstein:2016bsq} that an additional mode called the Glauber mode is necessary in such region for a consistent SCET,
and we expect that our construction of gravity SCET can be suitably modified for the Regge limit by adapting the framework of~\cite{Rothstein:2016bsq} for gravity.
As discussed in~\cite{Rothstein:2016bsq}, rapidity renormalization group~\cite{Chiu:2007yn, Chiu:2007dg, Chiu:2011qc, Chiu:2012ir} plays an important role. 
Since we have adopted the position-space formulation of SCET for our gravity SCET as opposed to the label SCET formalism as in~\cite{Rothstein:2016bsq}, 
it may be hoped that a simplification in dealing with rapidity renormalization group may be achieved by adapting the analytic regulator method (originally proposed by~\cite{Becher:2011dz} with a fully consistent perturbative treatment by~\cite{Jaiswal:2015nka}) for the Regge region and, most importantly, for curved spacetimes.
Especially, it must be checked whether a consistency of analytic regularization as shown for QCD SCET~\cite{Becher:2011dz} holds for gravity SCET or not, especially in the Regge region.

Within or away from our target phase space, 
the study of RPI in the presence of gravity is important and interesting~\cite{progress}.
As we pointed out above, such study should constitute an essential part of establishing soft and collinear theorems for gravity beyond the leading power in $\la$.
Especially, the concrete examples discussed in Appendices~\ref{a.Phi4}--\ref{a.ssff} hint at uncovered relations that relate the coefficients of operators of the type~(a) (see \Sec{other_lambdas}) to those of LP operators.
If this relation is due to RPI, 
that would mean that the RPI transformations in gravity SCET should depend on graviton fields.
This is not unexpected from the viewpoint that $n_i$-collinear RPI transformations are the $n_i$-collinear local Lorentz transformations that are ``broken'' by the introduction of $n_{\pms_i}$ and the choice of an $n_i$-collinear freely falling local inertial frame should depend on the $n_i$-collinear graviton field configuration.
Unlike in the case of QCD SCET, RPI in gravity SCET is an aspect of the factorized gauge symmetry, especially the Lorentz part of diff$\times$Lorentz, so it should inevitably involve the graviton field. 
It is also possible that the relations hinted at by the calculations in Appendices~\ref{a.Phi4}--\ref{a.ssff} are not due to RPI but instead are an indication of the existence of a ``universality class'' of gravity SCET\@.
Then, classifying all possible universality classes of gravity SCET would be an interesting problem.

Finally, 
as we alluded to in Introduction, it is interesting to study possible relations between infinite dimensional symmetries suggested by gravity SCET and those of the full theory discovered by~\cite{Strominger:2013jfa, He:2014laa, Kapec:2014opa}. 
In this regard, Ref.~\cite{Larkoski:2014bxa} already notes a connection between the RPI in QCD SCET and the M\"obius group.
We also noted that an EFT is ``half way'' between the full theory lagrangian and the $S$-matrix elements in that the path integrals are partly done by integrating out the modes outside the target phase space. 
So, it may be also possible to find an EFT that captures, at the lagrangian level, some of the amazing properties of the gravitational $S$-matrix such as ``gravity $=$ gauge$^2$''~\cite{Kawai:1985xq, Bern:2008qj}. In particular, it would be nice to find a logic of constructing an EFT that ``automatically'' leads to the double-copy symmetry structure discovered ingeniously by Ref.~\cite{Cheung:2016say} within the full thoery. 
We hope that our derivation of gravity SCET might serve as a useful prototype or guide for  further development of EFTs as a means to explore gravity amplitudes.

\vskip 1em
\noindent
{\bf Acknowledgment:}
The authors of this work are supported by the US Department of Energy 
under grant DE-SC0010102.

\appendix
\section{Appendices: Explicit Examples}
In the following appendices, we consider examples with four collinear sectors in three different full theories, and compare tree-level full-theory amplitudes with the corresponding SCET amplitudes at the LP and NLP\@.
For simplicity, we ignore the soft sector.
We will see how the collinear Lorentz and diff Wilson lines (\eq{LorentzWilson} and~\eq{diffWilson}) as well as the Ricci and Riemann tensors (\eq{Rimunu} and~\eq{Rimunurhosigma}) appear in effective SCET operators.
The examples below also demonstrate a practical power of gravity SCET\@. 
The amplitude of each example would require lengthy calculations with laborious expansions in $\la$ to the NLP followed by tricky cancellations of a large number of BLP and LP terms.
In contrast, the SCET directly gives us the NLP amplitude in terms of a few coefficients of effective operators, which also tells us which full-theory amplitude we should handpick to determine those few coefficients with the least effort.

\subsection{The $\phi^4$ full theory}
\label{a.Phi4}
Here, we consider a full theory with one massless real scalar $\phi$ coupled to gravity:
\beq
\cL = -\fr12 R + \fr 12 \bar{g}^{\mu\nu}\del_\mu\phi\,\del_\nu\phi - \fr{1}{4!}\ka\phi^4 
\,,\eql{Phi4:full}
\eeq
where $R$ is the Ricci scalar and we have set $\Mpl = 1$.
We will first consider a purely hard process $\phi_1 \phi_2 \to \phi_3 \phi_4$ (where $\phi_i$ is $n_i$-collinear by definition) at tree level and construct a corresponding purely hard SCET operator.
We will then consider a process $\phi_1 \phi_2 \to \phi_3 \phi_4 g_1$ where $g_1$ is an $n_1$-collinear graviton.
We will calculate the amplitude of this process at tree-level at the LP and NLP in the full theory, and see how that is reproduced from the LP and NLP pieces of the SCET lagrangian.

As described in \Sec{matching}, 
the first step is to construct $\cS_{\text{pure}} = \int \dd^4x \, \cL_\text{pure}$. 
For the process $\phi_1 \phi_2 \to \phi_3 \phi_4$, 
the form of $\cL_{\text{pure}}$ to be matched is 
\beq
\cL_\text{pure}(\phi_1, \ldots, \phi_4)
=&
\int\!
\dd s_{{}_1\!} \, \dd s_{{}_2\!} \, 
\dd s_{{}_3\!} \, \dd s_{{}_4\!} \, 
C(s_{{}_1}, s_{{}_2}, s_{{}_3}, s_{{}_4}) \>
\phi^\PD_{1\!} (x_1) \, 
\phi^\PD_{2\!} (x_2) \, 
\phi^\PD_{3\!} (x_3) \, 
\phi^\PD_{4\!} (x_4)
\,,\eql{Phi4:pure}
\eeq
where $\ds{x_i \equiv x + s_{{}_i\!} n_{\ms_i}}$.
Matching the amplitude from $\cL_\text{pure}$ to that from the full theory~\eq{Phi4:full} at tree level,
we get
\beq
C(s_1,s_2,s_3,s_4)=-\ka\, \de(s_1) \, \de(s_2) \, \de(s_3) \, \de(s_4)
\,.\eql{Phi4:C}
\eeq
(So, this is a special case where $\cL_\text{pure}$ happens to be local.)

Let us now turn on gravity and consider the process $\phi_1 \phi_2 \to \phi_3 \phi_4 g_1$ at tree level.
In the full theory, there are two sources of one-graviton couplings: 
$\sqrt{-g}=1 + h/2 + \cO(h^2)$ and $g^{\mu\nu}=\eta^{\mu\nu}-h^{\mu\nu} + \cO(h^2)$. 
Thus, the relevant interactions in the full theory are given by
\beq
\sqrt{-g} \, \cL_\text{int} 
= 
\fr{h}{2} \lt( \fr{1}{2}\del^\mu\phi\,\del_\mu\phi - \fr{1}{4!}\ka\phi^4 \rt)\! - \fr{1}{2}h^{\mu\nu}\del^\mu\phi\,\del_\nu\phi
\,.
\eeq
Therefore, in the process $\phi_1 \phi_2 \to \phi_3 \phi_4 g_1$,
the graviton $g_1$ can be emitted from any of $\phi_1$, \ldots, $\phi_4$ legs or from the $\phi^4$ vertex (see~\Fig{phi4}).
As discussed in \Sec{action_structure},
the amplitude of $g_1$ emitted from $\phi_1$ (\Fig{Phi4:phi1}) is the same in the full and effective theories, because it is a process occurring within the $n_1$-collinear sector alone.
Hence, we compare the full-theory amplitude from all the other diagrams with the SCET amplitude from $\cL_\text{hard}$.
In the full theory, the diagrams with $g_1$ emitted from the $\phi_{2,3,4}$ legs (\Fig{Phi4:phi2} and two similar diagrams with $g_1$ emitted from $\phi_{3,4}$) individually contain BLP and LP graviton couplings. 
But when we carefully expand the amplitudes in powers of $\la$ and add them together, 
all the BLP and LP terms cancel out and we are left with NLP contributions as we expect from the EFT\@.
The diagram with $g_1$ emitted from the $\phi^4$ vertex (\Fig{Phi4:vertex}) is already NLP by itself. 
At the end of the (very long) day, the full-theory amplitude for $\phi_1 \phi_2 \to \phi_3 \phi_4 g_1$ with $g_1$ emitted from $\phi_{2,3,4}$ or $\phi^4$ has the form $\I \cM_1 + \I \cM_2 + \I \cM_3$ with
\beq
\I\cM_i = -\I \ka \, \cA_i 
\qquad(i=1,2,3)
\eeq
where
\begin{align}
\cA_1
&=
\fr{p_1^\mu \, h_{\ms\mu}}{q_\ms} - \fr{(p_1 \dt\, q) \, h_{\ms\ms}}{2(q_\ms)^2} 
\,,\eql{Phi4DiffWL}\\
\cA_2
&=
\lt[ \fr{1}{(n_{\ps_1} \dt\, n_{\ps_3})(n_{\ms_3} \dt\, p_3^\PD)} 
\biggl(
\fr{p_3^{\tilde{\al}} p_3^{\tilde{\be}} h_{\tilde{\al} \tilde{\be}}}{2q_\ms} 
-\fr{(p_{3}^\pp \dt\, q^{\pp}) \, p_3^{\tilde{\al}} \, h_{\ms \tilde{\al}}}{(q_\ms)^2}
+\fr{(p_{3}^\pp \dt\, q^\pp)^2 \, h_{\ms\ms}}{2(q_\ms)^3} 
\biggr) \rt]
\eql{Phi4Riemann}\\
&\quad+\big[ 3\to4 \big] - \big[ 3\to2\big ]
\,,\nonumber\\
\cA_3
&=
\fr{1}{2} \lt( -\fr{h_{\tilde{\al}}^{\tilde{\al}}}{2} + \fr{q^{\tilde{\al}} h_{\ms \tilde{\al}}}{q_\ms} - \fr{(q_\pp \dt\, q_\pp) \, h_{\ms\ms}}{2(q_\ms)^2} \rt)
\,.\eql{Phi4Ricci}
\end{align}
In these expressions, all lightcone indices $-, +, \perp$ refer to the $n_1$-collinear coordinates, and similarly $h_{\mu\nu}$ is the $n_1$-collinear graviton field, 
treated as an external field and not necessarily on-shell. 
All the matter particles, on the other hand, are taken to be on-shell. 
A ``$\ds{\widetilde{\phantom{m}}}$'' on an index (such as $\tilde{\al}$) means that the index only refers to the $\perp$ components in the $n_1$-collinear coordinates.
The momenta of $\phi_{1,2,3,4}$ and $g_1$ are denoted by $p_{1,2,3,4}$ and $q$, respectively, where $p_{1,2}$ are ingoing while $p_{3,4}$ and $q$ outgoing. 
As noted above, every and each term in $\I\cM_{1,2,3}$ is NLP\@.

\begin{figure}[tbp]
\centering
\subfigure[]{
\includegraphics[scale=0.75]{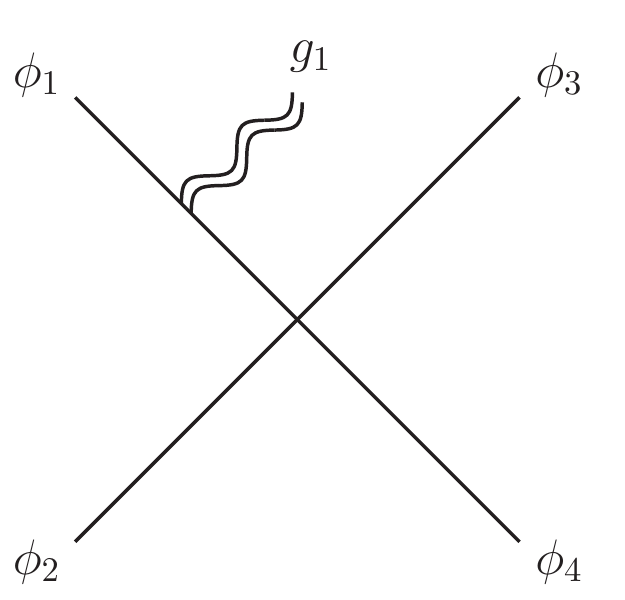}
\label{f.Phi4:phi1}%
}
\subfigure[]{
\includegraphics[scale=0.75]{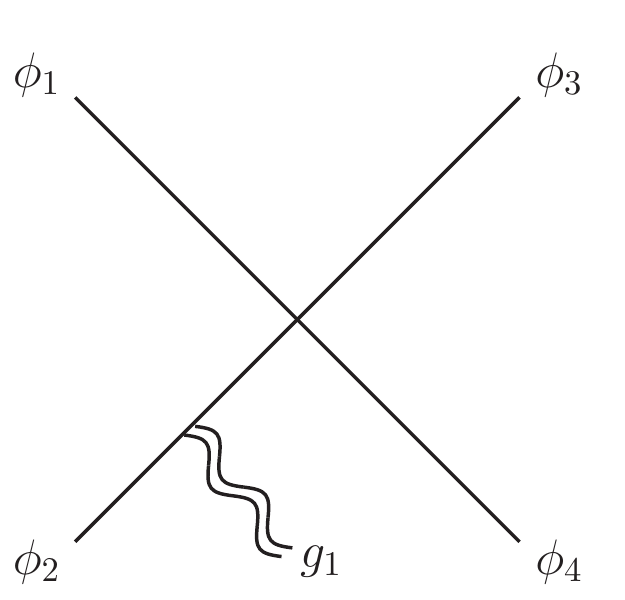}
\label{f.Phi4:phi2}%
}
\subfigure[]{
\includegraphics[scale=0.75]{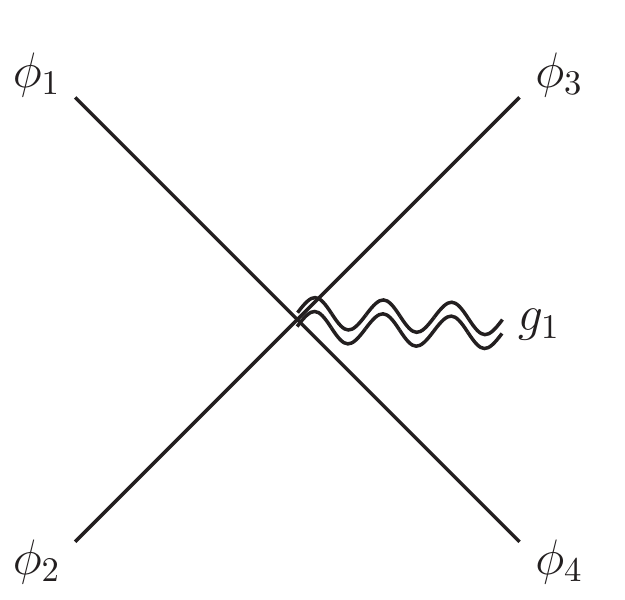}
\label{f.Phi4:vertex}%
}
\caption{\label{f.phi4} Representative full-theory diagrams in the $\phi^4$ theory of Appendix~\ref{a.Phi4}.}
\end{figure}

Let us now ask where $\I\cM_{1,2,3}$ come from in the SCET\@.
First, as described in \Sec{recipe},
each $\phi_i(x)$ in $\cL_\text{pure}$ should be replaced by 
$\ds{\phi'_i(x) \equiv V_i(x) \, \phi_i(x)}$,
where $V_i(x)$ is the $n_i$-collinear diff Wilson line~\eq{diffWilson}.
(Recall that we are ignoring the soft sector for simplicity here, so there's no $Y_i$.)
Let us refer to this part of $\cL_\text{hard}$ as $\cL_\text{hard-1}$, i.e., 
\beq
\cL_\text{hard-1}(\phi_1, \ldots, \phi_4, h) = \cL_\text{pure}(\phi'_1, \ldots, \phi'_4)
\,.\eql{Lhard-1}
\eeq
One can verify that the $n_1$-collinear graviton couplings from expanding $\cL_\text{hard-1}$ to the NLP exactly reproduce $\I\cM_1$.  

Next, the full-theory amplitude $\I\cM_2$ is reproduced in the SCET by an operator $\cL_\text{hard-2}$ containing an $n_1$-collinear Riemann tensor $R^1_{\mu\nu\rho\sg}$,%
\footnote{Our sign conventions for the curvature tensors are:
$R^{\rho}_{~\sg\mu\nu} = \del_\mu \Ga^\rho_{\nu\sg} - \del_\nu \Ga^\rho_{\mu\sg} + \ldots$ and $R_{\mu\nu} = R^\rho_{~\mu\rho\nu}$.}
where
\beq
&\cL_\text{hard-2} \\
=&
\int\! \dd s_{{}_1\!} \, \cdots \, \dd s_{{}_4\!} 
\int'\!\! \dd s_{{}_5\!} \fr{-s_{{}_5}^2}{2} \, 
C(s_{{}_1}, s_{{}_2}, s_{{}_3}, s_{{}_4}) \>
n_{\ms_1}^\al n_{\ms_1}^\be  R^1_{\al \mu \be \nu}(x + s_{{}_5\!} n_{\ms_1})
\,\times\\
&\,\times
\! \lt( \biggl[ 
\int_0^{\infty}\!\! \dd u \, 
\phi'_1(x_1) \, 
\phi'_2(x_2)\,\fr{\del^\mu \del^\nu \phi'_3(x_3 + u n_{\ms_3})}{(n_{\ps_1} \dt\, n_{\ps_3})} \,\phi'_4(x_4)  \biggr]\! +
[3\lr4] + [3\lr2, u \to -u] \rt) .
\eql{Phi4RiemannEFT}
\eeq
Of course, this expression needs to be expanded to the NLP\@.
In doing so, the $\mu$ and $\nu$ indices only need to be in the $\perp_1$ directions,
as we described below~\eq{Rimunurhosigma}.
Together with the $n_{\ms_1}^\al n_{\ms_1}^\be$, we will be picking up only the $R^1_{-\pp-\pp}$ component.
Since this is already $\cO(\la)$, 
all the $\phi'_i$'s in~\eq{Phi4RiemannEFT} should be replaced by the respective $\phi_i$'s. 

Finally, the full-theory amplitude $\I\cM_3$ is reproduced in the SCET by an operator $\cL_\text{hard-3}$ containing an $n_1$-collinear Ricci tensor $R^1_{\mu\nu}$,
where
\beq
&\cL_\text{hard-3} \\
=&
\int\! \dd s_{{}_1\!} \, \cdots \, \dd s_{{}_4\!} \int'\!\! \dd s_{{}_5\!} \,
\fr{s_{{}_5}}{2} \, C(s_{{}_1}, s_{{}_2}, s_{{}_3}, s_{{}_4}) \>
n_{\ms_1}^\al n_{\ms_1}^\be  R^1_{\al \be}(x + s_{{}_5\!} n_{\ms_1}) \,
\phi'_1(x_1) \, 
\phi'_2(x_2) \,
\phi'_3(x_3) \,
\phi'_4(x_4) .
\eql{Phi4RicciEFT}
\eeq
Again, since $n_{\ms_1}^\al n_{\ms_1}^\be  R^1_{\al \be}$ is already $\cO(\la)$,
all the $\phi'_i$'s should be replaced by the respective $\phi_i$'s.

\subsection{The $\phi^3$ full theory}
\label{a.Phi3}
Here we give another example, repeating the same process above but this time with a $\phi^3$ interaction in full theory. 
\beq
\cL = -\fr12 R + \fr 12 \bar{g}^{\mu\nu}\del_\mu\phi\,\del_\nu\phi - \fr{1}{3!}\ka \phi^3 
\,,\eql{Phi3:full}
\eeq
First, for $\phi_1 \phi_2 \to \phi_3 \phi_4$ without graviton emissions, 
the overall form of $\cL_\text{pure}$ remains as in~\eq{Phi4:pure}, but the Wilson coefficient $C$ in~\eq{Phi4:C} is now replaced by $C'$:
\beq
C'(s_1,s_2,s_3,s_4)
=\fr{\ka^2 \, \theta(-s_1)}{2}
\bigg[ 
\fr{\theta(-s_2) \, \de(s_3) \, \de(s_4)}{n_{\ps_1} \dt\, n_{\ps_2}} 
- \fr{\de(s_2) \, \theta(s_3) \, \de(s_4)}{n_{\ps_1} \dt\, n_{\ps_3}} 
- \fr{\de(s_2) \, \de(s_3) \, \theta(s_4)}{n_{\ps_1} \dt\, n_{\ps_4}}
\bigg]\,
\eql{CPhi3}
\eeq
where the three terms in $C'$ correspond to the $s$-, $t$-, and $u$-channel exchange diagrams of the full theory, respectively. 
We see that $\cL_\text{pure}$ displays a characteristic nonlocality of SCET, 
where the nonlocality comes from integrating out the highly off-shell virtual $\phi$ because such highly off-shell modes are not degrees of freedom of the SCET, unlike the collinear and soft modes.
The signs of the arguments of the step functions above are chosen to ensure the integration over the respective $s_i$ converges with the usual prescription of adding $\I\ep$ to the energy with a positive infinitesimal $\ep$.

Next, for $\phi_1 \phi_2 \to \phi_3 \phi_4 g_1$,
direct calculations show that
the full-theory amplitudes $\I\cM_{1,2,3}$ of \eq{Phi4DiffWL}--\eq{Phi4Ricci} are replaced in the $\phi^3$ theory by $\I\cM'_{1,2,3}$:
\beq
\I\cM'_1 = -\I\ka^2 \Big( \fr{1}{s_q} + \fr{1}{t_q} + \fr{1}{u_q} \Big) \cA_1
\,,\quad
\I\cM'_{2,3} = -\I\ka^2 \Big( \fr{1}{s_0} + \fr{1}{t_0} + \fr{1}{u_0} \Big) \cA_{2,3}
\eql{Phi3_amp}
\eeq
where $\cA_{1,2,3}$ are defined in~\eq{Phi4DiffWL}--\eq{Phi4Ricci}, 
and the Mandelstam variables are defined as 
$s_q = \ds{2 (p_1 \!-\! q) \dt p_2}$,
$t_q = \ds{-2 (p_1 \!-\! q) \dt p_3}$, 
and $u_q = \ds{-2 (p_1 \!-\! q) \dt p_4}$
with $s_0$, $t_0$, $u_0$ being $s_q$, $t_q$, $u_q$ with $q=0$, respectively.
One can verify that the full-theory amplitudes $\I\cM'_{1,2,3}$ are reproduced in the SCET by the operators~$\cL_\text{hard-$1,2,3$}$ in~\eq{Lhard-1}--\eq{Phi4RicciEFT} with $C$ replaced by $C'$,
where the dependencies on the Mandelstam variables in~\eq{Phi3_amp} all come from the step functions in $C'$. 

The fact that $C$ is just replaced by $C'$ is expected for the contributions from $\cL_\text{hard-1}$ because $\cL_\text{hard-1}$ is literally just $\cL_\text{pure}$ with each $\phi_i$ replaced by $\phi'_i = V_i \phi_i$.
It may be surprising for $\cL_\text{hard-$2,3$}$ in that those operators are gauge invariant by themselves so they are not related by gauge symmetry to $\cL_\text{pure}$.
We even have diagrams like \Fig{phi3} that do not have the $1/s + 1/t + 1/u$ structure.
There are two possible stories after this. 
One is that the $\phi^4$ and $\phi^3$ theories above belong to the same ``universality class'' of gravity SCET, so we only need to calculate one theory to fix the structures and coefficients of $\cL_\text{hard-$2,3$}$.
The other possibility is that there is a symmetry---perhaps RPI---that actually completely fixes the structures and coefficients of $\cL_\text{hard-$2,3$}$ once $\cL_\text{pure}$ is given.
Both possibilities are interesting and deserve a further dedicated study~\cite{progress}.

\begin{figure}[tbp]
\centering 
\includegraphics[scale=0.8]{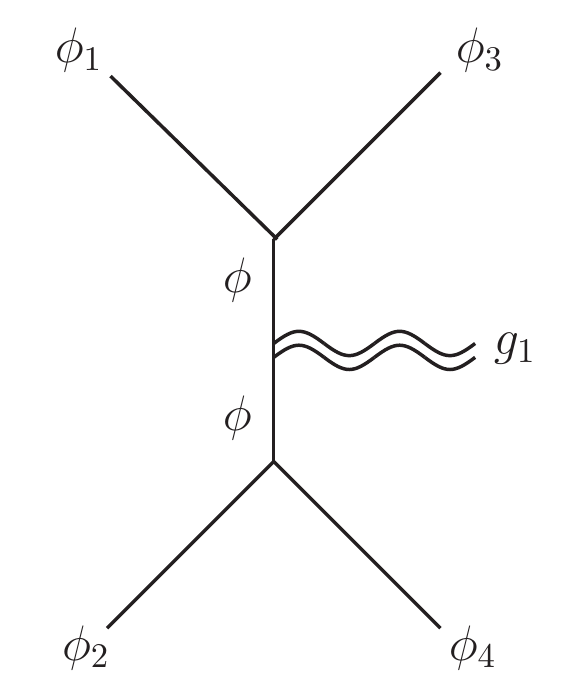}
\caption{\label{f.phi3} A full-theory diagram with the graviton emitted from the internal propagator in the $\phi^3$ theory.}
\end{figure}
%

\subsection{The $\phi\phi\psi\psi$ full theory}
\label{a.ssff}
Our third example is meant to pick up the collinear Lorentz Wilson line $W_r^i$ of~\eq{LorentzWilson}.
We also see RPI plays the role of type~(b) of \Sec{other_lambdas}.
Consider the full theory given by
\beq
\cL = 
-\fr{1}{2}R 
+\I \overline{\psi} \ga^\mu \DD_{\!\mu} \psi + \fr12 \bar{g}^{\mu\nu} \del_\mu\phi\,\del_\nu\phi - \fr{1}{2}\ka\overline{\psi}\psi\phi^2
\,,\label{e.ssffL}
\eeq
where the covariant derivative, $\DD_\mu$, is defined around~\eq{spin_connection}.
We then consider the processes $\psi_1 \phi_2 \to \psi_3 \phi_4$ and $\psi_1 \phi_2 \to \psi_3 \phi_4 g_1$ as we did for the $\phi^4$ and $\phi^3$ theories. 

First, for $\psi_1 \phi_2 \to \psi_3 \phi_4$ without a graviton emission, 
the form of $\cL_\text{pure}$ remains as in~\eq{Phi4:pure} 
except for the replacements $\phi_1 \to (P_1^+ \psi_1^\PD)_a$ and $\phi_3 \to (\wba{P_3^+ \psi_3^\PS})^a = (\wba{\psi}_3^\PD P_3^-)^a$ with the the helicity projection operators~\eq{hel-projection}, 
where $a$ is a Dirac spinor index that is summed over,
which leads to the spinor structure $\wba{\psi}_3(x_3) P_3^- P_1^+ \psi_1(x_1)$ in $\cL_\text{pure}$.
The coefficient $C$ remains as in~\eq{Phi4:C}.
 
Now consider $\psi_1 \phi_2 \to \psi_3 \phi_4 g_1$ in the full-theory.
The amplitudes $\I\cM_{1,2,3}$ of \eq{Phi4DiffWL}--\eq{Phi4Ricci} 
are now replaced with
\beq
\I\cM''_i
=
-\I \ka \, (\wba{\psi}_3^\PD P_3^- P_1^+ \psi_1^\PD) \cA_i
\qquad (i=1,2,3)
\eeq
with $\cA_{1,2,3}$ given by~\eq{Phi4DiffWL}--\eq{Phi4Ricci}.
Here, $\psi_{1,3}$ denote appropriate on-shell spinor wave functions, not the field operators, but there should be no confusion.
Most importantly, there is a new term in the amplitude:
\beq
\I\cM''_4
=& \fr{\I\ka}{8 q_\ms} \, \wba{\psi}_3^\PD P_3^- \Bigl(  
(q_\ms h_{\ps\ms} - q_\ps h_{\ms\ms}) \, [\ga^\ms, \ga^\ps] \,  P_1^+ \\
&\hspace{5em}
+ (q_{\tilde{\al}} h_{\tilde{\be} \ms}) \, [\ga^{\tilde{\al}}, \ga^{\tilde{\be}}] \, P_1^+ 
+ (q_{\tilde{\al}} h_{\ms\ms} - q_\ms h_{\tilde{\al} \ms}) \, [\ga^{\tilde{\al}}, \ga^\ms] \, P_1^- 
\Bigr) \psi_1^\PD
\,,\eql{e.ssffLorentzWL}
\eeq
where cancellations of BLP and LP contributions are already taken care of, so every term in this expression is NLP\@.

On the EFT side, $\cL_\text{hard-1}$ has the same form as~\eq{Lhard-1} except for the replacements $\phi'_1 \to (\psi'_1)_a$ and $\phi'_3 \to (\wba{\psi}'_3)^a$, where $a$ is a Dirac spinor index that is summed over, and $\psi'_i$ ($i=1,3$) now also include a Lorentz Wilson line:
\beq
\psi'_i(x) = V_i(x) \, W_{\text{\tiny D}}^i(x) \, \psi_i(x)
\,,
\eeq
where $W_{\text{\tiny D}}^i(x)$ is the $n_i$-collinear Lorentz Wilson line~\eq{LorentzWilson} for the Dirac spinor representation. 
Also, note the absence of helicity projection operators in the replacements $\phi'_1 \to (\psi'_1)_a$ and $\phi'_3 \to (\wba{\psi}'_3)^a$.
This is to take into account constraints from RPI (item~(b) of \Sec{other_lambdas}).
Then, one can verify that the terms coming from the diff Wilson lines in $\cL_\text{hard-1}$ reproduce the amplitude $\I\cM''_1$ as before. 
Most importantly, the terms from the $n_1$-collinear Lorentz Wilson line in $\cL_\text{hard-1}$
exactly reproduce $\I\cM''_4$.

Finally, the amplitudes $\I\cM_{2,3}$ are reproduced from $\cL_\text{hard-$2,3$}$ of~\eq{Phi4RiemannEFT} and~\eq{Phi4RicciEFT} with the same $C$ but with the replacements $\phi_1 \to (P_1^+ \psi_1^\PD)_a$ and $\phi_3 \to (\wba{\psi}_3^\PD P_3^-)^a$ with $a$ summed over.
Again, we observe the same clear pattern as we discussed in Appendix~\ref{a.Phi3}.


\providecommand{\href}[2]{#2}\begingroup\raggedright\endgroup

\end{document}